\documentclass[a4paper,11pt,fleqn]{article}

\usepackage{amsmath,amssymb,amsthm,enumerate,cite}%,backref

\newcommand{\todo}[1][\null]{\ensuremath{\clubsuit}}

\newcommand{\p}{\partial}
\newcommand{\const}{\mathop{\rm const}\nolimits}
\newcommand{\Equiv}{\mathop{ \sim}}

\newcommand{\sign}{\mathop{\rm sign}\nolimits}

\newcommand{\CV}{\mathop{\rm CV}\nolimits}
\newcommand{\CL}{\mathop{\rm CL}\nolimits}
\newcommand{\Ch}{\mathop{\rm Ch}\nolimits}

\newcounter{tbn}

\newcounter{mcasenum}
\renewcommand{\themcasenum}{{\rm\arabic{mcasenum}}}

\newtheorem{theorem}{Theorem}
\newtheorem{lemma}{Lemma}
\newtheorem{corollary}{Corollary}
\newtheorem{proposition}{Proposition}
\newtheorem*{proposition*}{Proposition}
{\theoremstyle{definition}
\newtheorem{definition}{Definition}
\newtheorem{example}{Example}
\newtheorem{note}{Note}
}

\begin{document}

\par\noindent {\LARGE\bf
Enhanced group analysis and conservation laws\\ 
of variable coefficient diffusion--reaction equations\\ 
with power nonlinearities
\par}

{\vspace{4mm}\par\noindent {\large 
O.~O.~Vaneeva$^{\dag 1}$, 
A. G. Johnpillai$^{\ddag 2}$, 
R.~O.~Popovych$^{\dag\natural 3}$ and 
C. Sophocleous$^{\S 4}$ }
\par\vspace{2mm}\par} {\vspace{2mm}\par\noindent {\it
${}^\dag$Institute of Mathematics of NAS of Ukraine, 3 Tereshchenkivska Str., Kyiv-4, 01601 Ukraine\\
$^\ddag$Department of Mathematics, Eastern University, Chenkalady, Sri Lanka  \\
$^\natural$Fakult\"at f\"ur Mathematik, Universit\"at Wien, Nordbergstra{\ss}e 15, A-1090 Wien, Austria\\
$^\S$Department of Mathematics and Statistics, University of Cyprus, Nicosia CY 1678, Cyprus\\[2mm]
}} {\noindent {\it $^1$vaneeva@imath.kiev.ua,
$^2$andrewgratienj@yahoo.com, $^3$rop@imath.kiev.ua,
$^4$christod@ucy.ac.cy\!}\par}

{\vspace{5mm}\par\noindent\hspace*{8mm}\parbox{144mm}{\small 
A class of variable coefficient (1+1)-dimensional nonlinear
diffusion--reaction equations of the general form
$f(x)u_t=(g(x)u^nu_x)_x+h(x)u^m$ is investigated. Different kinds
of equivalence groups are constructed including ones with
transformations which are nonlocal with respect to arbitrary
elements. For the class under consideration the complete group
classification is performed with respect to convenient equivalence
groups (\emph{generalized extended} and \emph{conditional} ones)
and with respect to the set of all point transformations.
Usage of different equivalences and coefficient gauges plays the
major role for simple and clear formulation of the final results.
The corresponding set of admissible transformations is described exhaustively.
Then, using the most direct method, we classify local conservation laws.
Some exact solutions are constructed by the classical Lie method.
}\par\vspace{4mm}}

%{\it Keywords}: Nonlinear diffusion equations; Equivalence transformations; Lie symmetries; Conservation laws

%\vspace{5mm}

\section{Introduction}

In this paper we study nonlinear partial differential equations
(PDEs) of the general class
\begin{equation} \label{eqRDfghPower}
f(x)u_t=(g(x)u^nu_x)_x+h(x)u^m
\end{equation}
with Lie symmetry point of view. Here $f=f(x)$, $g=g(x)$ and
$h=h(x)$  are arbitrary smooth functions of their variables,
$f(x)g(x)\neq0$, $n$ and $m$ are arbitrary constants. The linear
case is excluded from consideration as well-investigated. We
additionally assume the diffusion coefficient to be
nonlinear, i.e. $n\not=0$. The case $n=0$ is quite singular and
will be investigated separately.

Various simple transport models of electron temperature in a
confined plasma are reducible to equations of the form
(\ref{eqRDfghPower}) \cite{kamin&rosenau1982}. Here $u$ represents
the temperature, $f(x)$ the density and $g(x)$ the
density-dependent part of the thermal diffusion. The term
$h(x)u^m$ corresponds to the heat source. Furthermore a number of
special cases of the class (\ref{eqRDfghPower}) have been used to
model successfully problems in mathematical physics, chemistry and
biology
\cite{crank1979,murray2002,murray2003,peletier1981,touloukian1970}.

While there is no existing general theory for solving nonlinear PDEs, 
many special cases have
yielded to appropriate changes of variables. Point transformations
are the ones which are mostly used. These are transformations in
the space of the dependent and the independent variables of a PDE.
Probably the most useful point transformations of PDEs are those
which form a continuous Lie group of transformations, which leave
the equation invariant. Symmetries of this PDE are then revealed,
perhaps enabling new solutions to be found directly or via
similarity reductions. The classical method of finding Lie
symmetries is first to find infinitesimal transformations, with
the benefit of linearization, and then to extend these to groups
of finite transformations. This~method is easy to apply and well
established in the last few
years~\cite{Olver1986,ibragimov1999,fushchich1993,bluman1989}. 
A~large number of diffusion-type equations has been studied using
Lie group analysis. See for example,
\cite{Dorodnitsyn1982,Ivanova&Sophocleous2006,Popovych&Ivanova2004NVCDCEs}.

A goal in the present work is to classify the Lie
symmetries of the general class (\ref{eqRDfghPower}). The problem
of classification of Lie symmetries is dated back in the late
fifties, when Ovsiannikov \cite{Ovsiannikov1959} determined all 
the forms of the well-known nonlinear diffusion equation
\[
u_t=[f(u)u_x]_x
\]
that admit such symmetries. Since this latter work a number of
articles on the classification of Lie symmetries of diffusion type
equations appeared in the literature. See for example,
\cite{Basarab-Horwath&Lahno&Zhdanov2001,
Cherniga&Serov1998,Dorodnitsyn1979,Dorodnitsyn1982,Lahno&Spichak&Stognii2002,Nikitin2005,
Popovych&Ivanova2004NVCDCEs,Ivanova&Sophocleous2006}. The Lie
symmetries of (\ref{eqRDfghPower}) when $f=g=h=1$ can be found in
\cite{Dorodnitsyn1979,Dorodnitsyn1982} (see
also~\cite[Chapter~10]{Ibragimov1994V1}). In fact, Dorodnitsyn
\cite{Dorodnitsyn1979,Dorodnitsyn1982} carried out the
classification for the nonlinear diffusion equation with a source,
\begin{equation}\label{EqNCCDR}
u_t=[k(u)u_x]_x+q(u).
\end{equation}
Recently Popovych and Ivanova \cite{Popovych&Ivanova2004NVCDCEs}
presented the classification of Lie symmetries for the general
class of diffusion-convection equations
\[
f(x)u_t=[g(x)A(u)u_x]_x+B(u)u_x
\]
and later Ivanova and Sophocleous \cite{Ivanova&Sophocleous2006}
derived the Lie symmetries for the class
\begin{equation}\label{EqCDfgh}
f(x)u_t=[g(x)A(u)u_x]_x+h(x)B(u)u_x.
\end{equation}

Furthermore in the present work, using the most direct method
\cite{Popovych&Ivanova2004ConsLawsLanl}, we carry out two
classifications of local conservation laws up to equivalence
relations. Conservations laws of (\ref{eqRDfghPower}) when
$f=g=h=1$ can be found in \cite{Dorodnitsyn&Svirshchevskii1983}.

The presence of three functions in equation
(\ref{eqRDfghPower}) makes our task very difficult. However
knowledge of the equivalence transformations for class
(\ref{eqRDfghPower}) enable us to consider an equivalent, but
simpler form. For this reason, in the next section we construct different kinds of
equivalence groups for the class under consideration.
Extensive investigation and usage of equivalence transformations have the two-fold effect for
further group analysis of class (\ref{eqRDfghPower}):
reduction of the number of cases to be studied and simplification of form of both equations and
basis elements of their Lie algebras in these cases.
Moreover, presentation of classification results in a closed form becomes possible only after
application of generalizations of usual point equivalence group.

The group classification of the equations from class
(\ref{eqRDfghPower}) is carried out in
Section~\ref{SectionOnRDfghLieSymmetries}. It is shown that usage
of the \emph{generalized extended} equivalence group and
\emph{conditional} equivalence groups plays a crucial role in
solving this problem. Importance of true choice of gauge for
arbitrary elements with respect to equivalence transformations is
discussed and illustrated by examples. Let us note that similar
investigation on generalizations of equivalence group and
different possibilities in choice of gauges was first fulfilled
in~\cite{Ivanova&Popovych&Sophocleous2005,Ivanova&Popovych&Sophocleous2006}
for class~\eqref{EqCDfgh}.

In Section~\ref{SectionOnAddEqiuvTrans} we look for the additional equivalence transformations between obtained cases of symmetry extension,
which are inequivalent with respect to transformations from the adduced equivalence groups.
As a result, the problem of group classification in class~(\ref{eqRDfghPower}) with respect to the set of all point transformations
is solved in passing.
Connection between additional equivalence transformations and conditional equivalence groups is demonstrated.

Results of Section~\ref{SectionClassificationOfAdmTrans} are key
for this paper. The set of \emph{form-preserving
transformations}~\cite{Kingston&Sophocleous1998} (called also
\emph{admissible}~\cite{Popovych&Eshraghi2005} ones) of
class~(\ref{eqRDfghPower}) is described exhaustively. Roughly
speaking, any point transformation which links a pair of equations
from the class under consideration is called an admissible
transformation for this class. See \cite{Popovych&Eshraghi2005}
for rigorous definitions and problem statements. At the best of
our knowledge, first the set of admissible transformations was
described in nontrivial case, namely for a class of generalized
Burgers equations, by Kingston and Sophocleous
in~\cite{Kingston&Sophocleous1991}. Although the same problem has
been already solved for a number of different classes of
PDEs~\cite{Borovskikh2004,
Kingston&Sophocleous1998,Kingston&Sophocleous2001,Popovych&Eshraghi2005,Popovych&Ivanova&Eshraghi2004Gamma},
these results are not well-known.

The set of admissible transformations of class~(\ref{eqRDfghPower}) has an interesting structure.
The class~(\ref{eqRDfghPower}) can be presented as a union of normalized subclasses.
Some of the subclasses admit nontrivial extensions of the equivalence group,
which will be conditional equivalence groups for the whole class~(\ref{eqRDfghPower}).
Moreover, there are subclasses intersecting each other. In such cases interaction of the corresponding conditional equivalence groups
gives admissible transformations which cannot be interpreted in group terms. 
By-products of investigation of admissible transformations are, in particular,
exhaustive description of the additional equivalence transformations between symmetry extension cases and
derivation of all condition resulting in nontrivial conditional equivalence groups.

After presenting necessary notions and tools (Section~\ref{SectionOnBasicDefsOfCLs}),
in Section~\ref{SectionOnRDfghConsLaws} we classify local conservations laws of the equations from class (\ref{eqRDfghPower}).

Studying equations (\ref{eqRDfghPower}) in the framework of Lie
group analysis is completed in
Section~\ref{SectionOnRDfghSimilaritySolutions} with construction
of exact solutions for some equations from this class. We use
results of fulfilled group classification and apply the classical
Lie method. The main idea of the section is to show simplification
introduced in finding exact solutions with equivalence
transformations.

\section{Equivalence transformations and choice \\of investigated class} \label{SectionOnEquivTransOfEqRDfghPower}

In this section we derive the equivalence transformations of
class~\eqref{eqRDfghPower}. These transformations enable us to
reduce the class under consideration to a simpler form. We also
present the equivalence transformations of the reduced class
and a special subclass.

\begin{note}
The value of the arbitrary element~$m$ is undefined if $h=0$.
In this case, it is most convenient for one to assume $m$ equal to $n+1$.
Sometimes the value $m=1$ is also acceptable.
\end{note}

The usual equivalence group~$G^{\sim}$ of
class~\eqref{eqRDfghPower} is formed by the nondegenerate point
transformations in the space of~$(t,x,u,f,g,h,n,m)$, which are
projectible on the space of~$(t,x,u)$, i.e. they have the form
\begin{gather*}
(\tilde t,\tilde x,\tilde u)=(T^t,T^x,T^u)(t,x,u), \\[0.5ex]
(\tilde f,\tilde g,\tilde h,\tilde n,\tilde
m)=(T^f,T^g,T^h,T^n,T^m)(t,x,u,f,g,h,n,m),
\end{gather*}
and transform any equation from class~\eqref{eqRDfghPower} for the
function $u=u(t,x)$ with the arbitrary elements $(f,g,h,n,m)$ to
an equation from the same class for the function $\tilde u=\tilde
u(\tilde t,\tilde x)$ with the new arbitrary elements~$(\tilde
f,\tilde g,\tilde h,\tilde n,\tilde m)$.

\begin{theorem}
$G^{\sim}$ consists of the transformations
\[
\begin{array}{l}
\tilde t=\delta_1 t+\delta_2,\quad \tilde x=\varphi(x), \quad
\tilde u=\delta_3 u, \\[1ex]
\tilde f=\dfrac{\delta_0\delta_1}{\delta_3\varphi_x} f, \quad
\tilde g=\dfrac{\delta_0\varphi_x}{\delta_3^{n+1}}\, g, \quad
\tilde h=\dfrac{\delta_0}{\delta_3^m\varphi_x} h, \quad \tilde
n=n, \quad \tilde m=m,
\end{array}
\]
where $\delta_j$ $(j=\overline{0,3})$  are arbitrary constants, $\delta_0\delta_1\delta_3\not=0$,
$\varphi$ is an arbitrary smooth function of~$x$, $\varphi_x\not=0$.
\end{theorem}

It appears that class~\eqref{eqRDfghPower} admits other
equivalence transformations which do not belong to~$G^{\sim}$ and
form, together with usual equivalence transformations, a {\it
generalized extended equivalence group}. Restrictions on
transformations can be weakened in two directions. We admit the
transformations of the variables $t$, $x$ and $u$ can depend on
arbitrary elements (the prefix ``generalized''~\cite{Meleshko1994}), and this
dependence are not point necessarily and have to become point with
respect to $(t,x,u)$ after fixing values of arbitrary elements.
The explicit form of the new arbitrary elements~$(\tilde f,\tilde
g,\tilde h,\tilde n,\tilde m)$ is determined via
$(t,x,u,f,g,h,n,m)$ in some non-fixed (possibly, nonlocal) way
(the prefix ``extended''). We construct the complete (in this
sense) generalized extended equivalence group~$\hat G^{\sim}$ of
class~\eqref{eqRDfghPower}, using the direct method
\cite{Kingston&Sophocleous1998}.

\begin{theorem}
The generalized extended equivalence group~$\hat G^{\sim}$ of
class~\eqref{eqRDfghPower} is formed by the transformations
\[
\begin{array}{l}
\tilde t=\delta_1 t+\delta_2,\quad \tilde x=\varphi(x), \quad
\tilde u=\psi(x) u, \\[1ex]
\tilde f=\dfrac{\delta_0\delta_1}{\varphi_x\psi^{n+2}} f, \quad
\tilde g=\dfrac{\delta_0\varphi_x}{\psi^{2n+2}}\, g, \quad
\tilde h=\dfrac{\delta_0}{\varphi_x\psi^{m+n+1}} h, \quad
\tilde n=n, \quad \tilde m=m,
\end{array}
\]
where $\delta_j$ $(j=0,1,2)$ are arbitrary constants, $\delta_0\delta_1\not=0$,
$\varphi$ is an arbitrary smooth function of~$x$, $\varphi_x\not=0$.
The function $\psi(x)$ is determined by the formula
\[
\psi(x)=\begin{cases} \bigl(1-(n+1)F(x)\bigr)^{-\frac 1{n+1}}, &
n\neq -1 \\ e^{F(x)}, & n=-1\end{cases}, \quad\mbox{where}\quad
F(x)=\delta_3\int \frac{dx}{g(x)}+\delta_4.
\]
\end{theorem}
\begin{note}
The above representation of the function $\psi(x)$ guarantees that
this family of transformations is continuously parameterized with
the parameter $n$.
\end{note}

From theorem 1 we deduce that the transformation
\begin{equation}\label{EqGaugeTransformationForG1}
\tilde t=t,\quad  \tilde x=\int \frac{dx}{g(x)},\quad \tilde u=u
\end{equation}
maps equation~(\ref{eqRDfghPower}) to
\[
\tilde f(\tilde x)\tilde u_{\tilde t}= (\tilde u^n \tilde
u_{\tilde x})_{\tilde x} + \tilde h(\tilde x)\tilde u^m,
\]
where $\tilde f(\tilde x)=g(x)f(x)$, $\tilde g(\tilde x)=1$ and
$\tilde h(\tilde x)=g(x)h(x)$. (Likewise any equation of
form~(\ref{eqRDfghPower}) can be reduced to the same form with
$\tilde f(\tilde x)=1$.) That is why,
without loss of generality, we can restrict ourselves to
investigation of the equations of the general form
\begin{equation} \label{eqRDfghPowerG1}
f(x)u_t=(u^nu_x)_x + h(x)u^m.
\end{equation}
All results on symmetries, solutions and conservation laws of class~(\ref{eqRDfghPowerG1}) can be extended to
class~(\ref{eqRDfghPower}) with transformations~\eqref{EqGaugeTransformationForG1}.

It appears that we can deduce the equivalence group for class~(\ref{eqRDfghPowerG1})
from Theorems~1 and~2 by setting $\tilde g=g=1$.
The results are summarized in the following theorem.

\begin{theorem}\label{TheoremOnEquivGroupg1}
The generalized equivalence group~$G^{\sim}_1$ of the
class~\eqref{eqRDfghPowerG1}, where $n\neq-1$, consists of the
transformations
\[
\begin{array}{l}
\tilde t=\delta_1 t+\delta_2,\quad \tilde x=\dfrac{\delta_3
x+\delta_4}{\delta_5 x+\delta_6}=:\varphi(x), \quad
\tilde u=\delta_7{\varphi_x}^\frac1{2n+2} u, \\[2ex]
\tilde f=\delta_1{\delta_7}^n{\varphi_x}^{-\frac{3n+4}{2n+2}} f,
\quad \tilde
h={\delta_7}^{-m+n+1}{\varphi_x}^{-\frac{m+3n+3}{2n+2}} h, \quad
\tilde n=n, \quad \tilde m=m,
\end{array}
\]
where $\delta_j$ $(j=\overline{1,7})$ are arbitrary constants,
$\delta_1\delta_7\not=0$ and $\delta_3\delta_6-\delta_4\delta_5=\pm1$.

For $n=-1$ transformations from group~$G^{\sim}_1$ take the form
\[
\begin{array}{l}
\tilde t=\delta_1 t+\delta_2,\quad \tilde x=\delta_3 x+\delta_4,
\quad
\tilde u=\delta_5e^{\delta_6 x} u, \\[1ex]
\tilde f=\dfrac{\delta_1}{{\delta_3}^2\delta_5e^{\delta_6 x}}  f,
\quad \tilde h=\dfrac{1}{{\delta_3}^2{\delta_5}^m e^{m\delta_6 x}}
h, \quad \tilde n=n, \quad \tilde m=m,
\end{array}
\]
where $\delta_j$ $(j=\overline{1,6})$ are arbitrary constants, $\delta_1\delta_3\delta_5\not=0$.
\end{theorem}

These equivalence transformations can be employed to simplify the
forms of $f(x)$ and $h(x)$ in the study of Lie symmetries of
(\ref{eqRDfghPowerG1}) which takes place in the next section.

\begin{note}
Theorem~\ref{TheoremOnEquivGroupg1} can be reformulated in the following way:
The usual equivalence group~$G^{\sim}_1$ of class~\eqref{eqRDfghPowerG1}
for values $n\not=-1$ is formed by the transformations
\[
\begin{array}{l}
\tilde t=\delta_1 t+\delta_2,\quad \tilde
x=\dfrac{1}{\varepsilon(n+1)\delta_3(1-(n+1)(\delta_3x+\delta_4))}-\dfrac{1}
{\varepsilon(n+1)\delta_3}=:\varphi(x),\\[2,5ex]
\tilde u=(1-(n+1)(\delta_3x+\delta_4))^{-\frac1{n+1}}
u={\varepsilon}^{\frac 1{2n+2}}{\varphi_x}^\frac1{2n+2} u
=: \psi(x) u, \\[1,5ex]
\tilde f=\delta_1{\varepsilon}^{\frac
n{2n+2}}{\varphi_x}^{-\frac{3n+4}{2n+2}} f, \quad \tilde
h={\varepsilon}^{\frac
{-m+n+1}{2n+2}}{\varphi_x}^{-\frac{m+3n+3}{2n+2}} h, \quad \tilde
n=n, \quad \tilde m=m,
\end{array}
\]
where $\delta_j$ $(j=\overline{1,4})$ and $\varepsilon$ are
arbitrary constants, $\varepsilon\delta_1\delta_3\not=0$.

This representation of the transformations from~$G^{\sim}_1$ is
more clumsily than in the theorem but it shows that the family of
the above transformations is continuously parameterized with the
parameter $n$, including the value $n=-1$. In fact,
\begin{displaymath}
\lim_{n \to -1} \varphi(x)= \frac {x}{\varepsilon}+\frac
{\delta_4}{\varepsilon\delta_3},\quad \lim_{n \to -1} \psi(x)=
e^{\delta_3x+\delta_4}.
\end{displaymath}
\end{note}

Since the parameters~$n$ and~$m$ are invariants of all the above equivalence transformations,
class~\eqref{eqRDfghPower} (or class~\eqref{eqRDfghPowerG1}) can be presented as the union of disjoint subclasses where
each from the subclasses corresponds to fixed values of~$n$ and~$m$.
This representation allow us to give other interpretation of the above results.
For example, the generalized equivalence group~$G^{\sim}_1$ from Theorem~\ref{TheoremOnEquivGroupg1},
more exactly its projection to the complementary set of variables and arbitrary elements, can be considered as
a family of usual equivalence groups of the subclasses parameterized with~$n$ and~$m$.

The question occurs: could one obtain wider equivalence groups for
some of the subclasses, apriori assuming the parameters~$n$
and~$m$ satisfy a condition? The answer is positive in the case of
$m=n+1$. Namely, the following statement is true.

\begin{theorem}\label{TheoremOnEquivGroupMN1}
The class of equations
\begin{equation}\label{eqRDfghPowerMN1}
f(x)u_t=(g(x)u^nu_x)_x+h(x)u^{n+1}
\end{equation}
admits the equivalence group~$G^{\sim}_{m=n+1}$ consisting of the transformations:
\begin{gather*}
\tilde t=\delta_1 t+\delta_2,\quad \tilde x=\varphi(x),\quad\tilde u=\psi(x)u,  \\
\tilde f=\frac{\delta_0\delta_1}{\psi^{n+2}\varphi_x}f,\quad
\tilde g=\frac{\delta_0\varphi_x}{\psi^{2n+2}}g,\quad \tilde
h=\delta_0\frac{h-\psi^{n+1}(\psi^{-(n+2)}\psi_xg)_x}{\psi^{2n+2}\varphi_x},\quad
(\tilde n=n),
\end{gather*}
where~$\varphi$ and~$\psi$ are arbitrary functions of~$x$,
$\delta_j$ $(j=0,1,2)$ are arbitrary constants, $\delta_0\delta_1
\varphi_x\psi \not=0$.
\end{theorem}

It should be emphasized that $G^{\sim}_{m=n+1}$ is the usual equivalence group of class~\eqref{eqRDfghPowerMN1} even through
$n$ is assumed as an arbitrary element.
Moreover, it is wider than the generalized extended equivalence group~$\hat G^{\sim}$ of whole class~\eqref{eqRDfghPower}.
We will also called $G^{\sim}_{m=n+1}$ as the \emph{conditional equivalence group} of class~\eqref{eqRDfghPower}
under the condition $m=n+1$ on the arbitrary elements.
It is a nontrivial conditional equivalence group of class~\eqref{eqRDfghPower} in the sense that
it is not a subgroup of~$\hat G^{\sim}$.
Let us note that the notion of conditional equivalence group was first used in classification of systems of two nonlinear Laplace equations
\cite{Popovych&Cherniha2001} (see also \cite{Popovych&Eshraghi2005}).

Gauging~$g$ with the condition~$g=1$, we impose the restrictions $\delta_0\varphi_x=\psi^{2n+2}$ on parameters of~$G^{\sim}_{m=n+1}$
and obtain the equivalence group $G^{\sim}_{1,m=n+1}$ of the class of equations
\begin{equation}\label{eqRDfghPowerG1MN1}
f(x)u_t=(u^nu_x)_x+h(x)u^{n+1}.
\end{equation}

\begin{corollary}\label{CorollaryOnEquivGroupMN1}
The equivalence group~$G^{\sim}_{1, m=n+1}$ of class~\eqref{eqRDfghPowerG1MN1}
is formed by the transformations
\begin{gather*}
\tilde t=\delta_1 t+\delta_2,\quad \tilde x=\varphi(x),\quad\tilde u=\psi(x)u,  \\
\tilde f=\frac{\delta_0^2\delta_1}{\psi^{3n+4}}f,\quad \tilde
h=\delta_0^2\frac{h-\psi^{n+1}[\psi^{-(n+2)}\psi_x]_x}{\psi^{4n+4}},\quad
(\tilde n=n),
\end{gather*}
where~$\varphi$ and~$\psi$ are arbitrary functions satisfying the
condition $\delta_0\varphi_x=\psi^{2n+2}$, $\delta_j$ $(j=0,1,2)$
are arbitrary constants, $\delta_0\delta_1\psi \not=0$.
\end{corollary}

Similarly to~$G^{\sim}_1$, $G^{\sim}_{1, m=n+1}$ is the generalized equivalence group for the whole class~\eqref{eqRDfghPowerG1MN1} 
since transformations of variables explicitly depend on the arbitrary element~$n$, 
and $G^{\sim}_{1, m=n+1}$ becomes the usual equivalence group under restriction to any subclass with a fixed value of~$n$.

Simultaneous usage of both the generalized equivalence group~$G^{\sim}_1$ and
the conditional equivalence group~$G^{\sim}_{1, m=n+1}$ via gauging arbitrary elements
leads to crucial simplification of solving the group classification problem and presentation of the obtained results.
See Note~\ref{NoteOnGauges} and Example~\ref{ExampleOnGaugesIfMN1} for some explanations.

There exists another condition on arbitrary elements, which gives a nontrivial conditional equivalence group.
All such conditions and the corresponding conditional equivalence groups are systematically investigated as components
of classification of admissible transformations in Section~\ref{SectionClassificationOfAdmTrans}.

\begin{note}
Due to physical sense of equation~~\eqref{eqRDfghPower}, the function~$u$ should satisfy the condition~$u\ge0$.
In this case we have to demand for the multipliers of~$u$ to be positive in all transformations.
If we avoid positiveness of~$u$ then we have to use the modular of~$u$ as base of powers which are not determined for negative values of base.
The same statement is true for similar expressions in transformations and other places.
The necessary changes in formulas are obvious.
\end{note}

\section{Lie symmetries}\label{SectionOnRDfghLieSymmetries}

In the section we present a complete classification of Lie symmetries for class~(\ref{eqRDfghPowerG1}).
We search for operators of the form $\Gamma=\tau(t,x,u)\partial_t+\xi(t,x,u)\partial_x+\eta(t,x,u)\partial_u$,
which generate one-parameter groups of point symmetry transformations of equations from class~(\ref{eqRDfghPowerG1}).
These operators satisfy the necessary and sufficient criterion of infinitesimal invariance, i.e.
action of the $r$-th prolongation~$\Gamma^{(r)}$ of $\Gamma$ to the ($r$-th order) DE
results in identically zero, modulo the DE under consideration. Here we require that
\begin{equation}\label{c1}
\Gamma^{(2)}\{f(x)u_t-u^nu_{xx} -nu^{n-1}u_x^2 - h(x)u^m\}=0
\end{equation}
identically, modulo equation~(\ref{eqRDfghPowerG1}).

After elimination of $u_t$ due to (\ref{eqRDfghPowerG1}),
equation~(\ref{c1}) becomes an identity in six variables, $t$, $x$, $u$, $u_x$, $u_{xx}$ and $u_{tx}$.
In fact, equation~(\ref{c1}) is a
multivariable polynomial in the variables $u_x$, $u_{xx}$ and $u_{tx}$. The
coefficients of the different powers of these variables must be
zero, giving the determining equations on the coefficients $\tau$, $\xi$ and $\eta$.
Since equation~(\ref{eqRDfghPowerG1}) has a specific form
(it is a quasi-linear evolution equation,
the right hand side of~(\ref{eqRDfghPowerG1}) is a
polynomial in the pure derivatives of $u$ with respect to $x$ etc),
the forms of the coefficients can be simplified.
That is, $\tau=\tau(t),$ $\xi=\xi(t,x)$~\cite{Kingston&Sophocleous1998} and, moreover,
$\eta=\zeta(t,x)u$ and $\xi=\xi(x)$ (see Section~\ref{SectionClassificationOfAdmTrans}).
To derive the latter condition, we partially split with respect to~$u$.
Finally we obtain the \emph{classifying} equations which include
both the residuary uncertainties in coefficients of the operator
and the arbitrary elements of the class under consideration:
\begin{gather*}
f_x\xi=f(n\zeta+\tau_t-2\xi_x), \qquad
2(n+1)\zeta_x=\xi_{xx}, \\[1ex]
uf^2\zeta_t-u^{n+1}f\zeta_{xx}+u^m(hf_x\xi-fh_x\xi+(1-m)fh\zeta-fh\tau_t)=0.
\end{gather*}
The third equation is a polynomial in~$u$ and should be split with
respect to~$u$ for all values of the parameters~$n$ and~$m$, which
correspond to the nonlinear case $n\not=0$. Splitting gives
essentially different results in three exclusive cases: 1. $m \ne
1, n+1$; 2. $m=1$; 3. $m=n+1$. The obtained equations enable us to
derive the forms of $\tau(t)$, $\xi(x)$, $\zeta(t,x)$, $f(x)$ and
$h(x)$ depending on values of~$n$ and~$m$ and consequently the
desired Lie symmetries will be constructed.

In Table~1 we list tuples of parameter-functions~$f(x)$ and $h(x)$, the constant parameters $m$ and $n$
and bases of the invariance algebras
in all possible inequivalent cases of Lie symmetry extension.
The operators from Table~1 form bases of the
maximal Lie invariance algebras iff the corresponding values of the parameters
are inequivalent to ones with more abundant Lie invariance algebras.
In Cases~2 and~5 we do not try to use equivalence transformations as much as possible
since otherwise a number of similar simplified cases would be derived,
see remarks after the table.

\begin{note}
It should be emphasized that we adduce only the cases of extensions of maximal Lie invariance algebra,
which are inequivalent with respect to~$G^{\sim}_1$ if $m\not=n+1$
and with respect to~$G^{\sim}_{1,m=n+1}$ if $m=n+1$ or $h=0$.
During solving the group classification problem, we derive some cases of extensions,
which are transformed with equivalence transformations to
representatives from the table in a nontrivial way.
Below we give some examples of such transformations.
\end{note}

\begin{example}
The `variable coefficient' equation
\begin{equation}\label{vbleequation1}
|x|^{-\frac{3n+4}{n+1}} u_t=(u^n u_x)_x+ \varepsilon |x|^{-\frac{3n+3+m}{n+1}}u^m, \quad
n\not=-1, \quad m\not=n+1,
\end{equation}
is transformed to the `constant coefficient' equation
\begin{equation}\label{constequation1}
\tilde u_{\tilde t}=({\tilde u}^n {\tilde u}_{\tilde x})_{\tilde x}+ \varepsilon {\tilde u}^m
\end{equation}
(Case~3 of Table~1 if $m\not=1$ and Case~6 if $m=1$) by the transformation
\[
\tilde t=t,\quad \tilde x=\frac1x=:\varphi(x), \quad
\tilde u=|x|^{-\frac1{n+1}} u=|{\varphi_x}|^\frac1{2n+2} u
\]
which belongs to the equivalence group~$G^{\sim}_1$ of the class~(\ref{eqRDfghPowerG1}).
Hence, using the maximal Lie invariance algebra of Case~3,
we can derive the basis elements of the maximal Lie invariance algebra of~(\ref{vbleequation1}):
\[
\partial_t,\quad
(n+1)x^2\partial_x+xu\partial_u,\quad
2(1-m)t\partial_t-(n+1-m)x\partial_x+\frac{n+1+m}{n+1}u\partial_u %\quad\mbox{if}\quad m\not=1
\]
if $m\not=1$ and
\[
\partial_t, \quad e^{-\varepsilon nt}(\partial_t+\lambda u\partial_u),\quad
(n+1)x^2\partial_x+xu\partial_u,\quad  n(n+1)x\partial_x-(n+2)u\partial_u %\quad\mbox{if}\quad m=1.
\]
if $m=1$. In the latter case we additionally assume $n\not=-4/3$ for the algebra to be really maximal.
\end{example}

\begin{example}
In an analogous way, the `variable coefficient' equation
\begin{equation}\label{vbleequation2}
e^x u_t=\left(\frac{u_x}u\right)_x+ \varepsilon e^x u
\end{equation}
is reduced by the transformation
$\tilde t=t,\ \tilde x=x, \ \tilde u=e^x u$
from the equivalence group~$G^{\sim}_1$ to the equation~\eqref{constequation1} with $n=-1$ and $m=1$ (Case~6 with $n=-1$).
The basis elements of the maximal Lie invariance algebra of~(\ref{vbleequation2}) have the form
\[
\partial_t, \quad e^{\varepsilon t}(\partial_t+\varepsilon u\partial_u),\quad
\partial_x-u\partial_u,\quad x\partial_x-(x+2)u\partial_u.
\]
\end{example}

\begin{table}\footnotesize
\renewcommand{\arraystretch}{1.6}
%\caption{Lie symmetries}
\begin{center}
\textbf{Table 1.} Results of group classification of class~\eqref{eqRDfghPower}
\\[2ex]
\begin{tabular}{|c|c|c|c|l|}
\hline
&$n$&$f(x)$&$h(x)$&\hfil Basis of $A^{max}$ \\
\hline
\multicolumn{5}{|c|}{General case}\\
\hline 1&$\forall$&$\forall$& $\forall$&
$\partial_t$\\
\hline 2&$\forall$&$f_1(x)$& $h_1(x)$&
$\partial_t, (d+2b-pn)t\partial_t+((n+1)ax^2+bx+c)\partial_x+(ax+p)u\partial_u$\\
\hline
3&$\forall$&1&$\varepsilon$&$\partial_t, \partial_x, 2(1-m)t\partial_t+(1+n-m)x\partial_x+2u\partial_u$ \\
\hline
\multicolumn{5}{|c|}{$m=1$, \quad $h\neq0$, \quad $(h/f)_x=0$}\\
\hline
4&$\forall$&$\forall$&$\varepsilon f$&$\partial_t, e^{-\varepsilon nt}(\partial_t+\varepsilon u\partial_u)$ \\
\hline 5&$\forall$&$f_1(x)$&$\varepsilon f$&$\partial_t,
e^{-\varepsilon nt}(\partial_t+\varepsilon u\partial_u), n((n+1)ax^2+bx+c)\partial_x+(nax+2b+d)u\partial_u$ \\
\hline 6&$\neq -\frac 43$&1&$\varepsilon$&$\partial_t, \partial_x,
e^{-\varepsilon nt}(\partial_t+\varepsilon u\partial_u), nx\partial_x+2u\partial_u$ \\
\hline 7&$-\frac 43$&1&$\varepsilon$&$\partial_t, \partial_x,
e^{\frac 43\varepsilon t}(\partial_t+\varepsilon u\partial_u),
-\frac 43 x\partial_x+2u\partial_u,
-\frac 13x^2\partial_x+xu\partial_u$ \\
\hline
\multicolumn{5}{|c|}{$m=n+1$ \quad or \quad $h=0$}\\
\hline
8&$\forall$&$\forall$&$\forall$&$\partial_t, nt\partial_t-u\partial_u$ \\
\hline 9&$\neq -\frac 43$&1&$\alpha x^{-2}$&$\partial_t,
nt\partial_t-u\partial_u, 2t\partial_t+x\partial_x$  \\
\hline 10&$\neq -\frac 43$&1&$\varepsilon$&$\partial_t,
nt\partial_t-u\partial_u, \partial_x$  \\
\hline
11&$\neq -\frac 43$&1&0&$\partial_t, \partial_x, nt\partial_t-u\partial_u, 2t\partial_t+x\partial_x$ \\
\hline 12&$-\frac 43$&$e^x$&$\alpha$&$\partial_t,
t\partial_t+\frac34u\partial_u, \partial_x-\frac34u\partial_u$  \\
\hline 13&$-\frac 43$&1&0&$\partial_t, \partial_x,
\frac 43 t\partial_t+u\partial_u, 2t\partial_t+x\partial_x, -\frac{1}{3}x^2\partial_x+xu\partial_u$ \\
\hline
\end{tabular}
\end{center}
Here  $\alpha$ is an arbitrary constant, $\alpha\not=0$ in Case~9, $\varepsilon=\pm1$,
\[
f_1(x)={\rm exp}\left [\int \frac{-(3n+4)ax+d}{(n+1)ax^2+bx+c}dx\right ],\quad
h_1(x)={\rm exp}\left [\int \frac{-(3(n+1)+m)ax+(n-m+1)p-2b}{(n+1)ax^2+bx+c}dx\right ],
\]
and it can be assumed up to $G^{\sim}_1$-equivalence that, if $n\neq-1$,
the parameter tuple~$(a,b,c,d,p)$ takes only the following inequivalent values:
\[
\{(0,1,0,\bar d,(\bar q+2)/(n+m-1)),\ (0,0,1,1,\check p),\ (0,0,1,0,1),\ (1/(n+1),0,1,\hat d,\hat p)\},
\]
where $\check p$ is an arbitrary constant; 
$\hat d\geqslant0$ and, if $\hat d=0$, $\hat p\geqslant0$;
\[
(\bar d,\bar q)\ne(0,0),\left(-\frac{3n+4}{n+1},-3-\frac m{n+1}\right);\quad
\bar d\geqslant-\frac{3n+4}{2(n+1)}\ \mbox{and, if\ } 
\bar d=-\frac{3n+4}{2(n+1)},\  \bar q\geqslant-\frac32-\frac m{2(n+1)}. 
\]
If $n=-1$, up to $G^{\sim}_1$-equivalence
the parameter tuple~$(a,b,c,d,p)$ can be assumed to belong to the set
\[
\{(0,1,0,d',p'),\ (0,0,1,0,1),\ (\varepsilon,0,1,0,p'')\},
\]
where
$d'$  and $p'$ are arbitrary constants, $p''\geqslant0$.
In Case~5 the parameter~$p$ should be neglected.
In Case~8 the parameter-functions~$f$ and~$h$ can be additionally
gauged with equivalence transformations from $G^{\sim}_{1,m=n+1}$.
For example, we can put $f=1$ if $n\not=-4/3$ and $f=e^x$
otherwise.
\end{table}

\begin{note}\label{NoteOnGauges}
Right choice of a gauge of the parameter tuple~$(f,g,h)$ is
another crucial point of our investigations. Really, the major
choice have been made from the very outset of classification when
the parameter-function~$g$ was put equal to 1. It is the gauge
that leads to maximal simplification of both the whole solving and
the final results. Although all gauges are equivalent from an
abstract point of view, only the gauge~$g=1$ allows one to
exhaustively solve the problem of group classification of
class~\eqref{eqRDfghPower} with reasonable quantity of
calculations. Even after the other simplest gauge~$f=1$ chosen,
calculations become too cumbersome and sophisticated.

The expression of transformations from the conditional equivalence group~$G^{\sim}_{m=n+1}$,
i.e. the equivalence group of the subclass~\eqref{eqRDfghPowerMN1} separated from class~\eqref{eqRDfghPower}
with the condition~$m=n+1$,
contains one more arbitrary function of~$x$ in comparison with the equivalence group of the whole class.
This fact makes an additional gauge of the parameter tuple~$(f,g,h)$ is possible and necessary in case of~$m=n+1$.
Different ways of the additional gauge with the conditional equivalence group~$G^{\sim}_{m=n+1}$ (or $G^{\sim}_{1,m=n+1}$)
are discussed in the next example.
\end{note}

\begin{example}\label{ExampleOnGaugesIfMN1}
Let $m=n+1$ and the gauge $g=1$ be already fixed.
The kernel $A^{\rm ker}_{1,m=n+1}$ of maximal Lie invariance algebras of the subclass~\eqref{eqRDfghPowerG1MN1}
is generated by the operators $\partial_t$ and $nt\partial_t-u\partial_u$ (Case~8 of Table~1).

It seems on the face of it that the optimal choice for the additional gauge is $h=0$.
This gauge results to essential simplification of both the general form of initial equations
and the determining equations. Indeed, the system of determining equations are then reduced to
the following one:
\[
\xi f_x=(n\zeta-2\xi_x+\tau_t)f, \quad \xi_{xx}=2(n+1)\zeta_x, \quad \zeta_{xx}=0,  \quad \zeta_t=0.
\]
Hence, $\zeta=C_1x+C_2$, $\xi=C_1(n+1)x^2+C_3x+C_4$, $\tau=C_5t+C_6$.
The main difficulty arises after taking into account the first equation.
It implies that extension of $A^{\rm ker}_{m=n+1}$ is possible iff
\[
(\alpha x^2+\beta x+\gamma)f_x=\delta f,
\]
where $\alpha$, $\beta$, $\gamma$ and $\delta$ are arbitrary constants, $(\alpha,\beta,\gamma)\not=(0,0,0)$.
Moreover, the equivalence group of the gauged class `$g=1$, $h=0$' is formed by the transformations
from~$G^{\sim}_{m=n+1}$ which additionally satisfy
the conditions $\delta_0\varphi_x=\psi^{2n+2}$, $(\psi^{-(n+1)}\psi_x)_x=0$.
Although classification may be carried out in this way,
complexity of both the expression of~$f$ in some extension cases and
equivalence transformations to be applied results in necessity of careful and sophisticated investigations
of reducibility between different cases.

Another way is to begin with gauging the parameter-function~$f$. The obtained gauge
\[
f=1 \quad\mbox{if}\quad n\not=-\frac43 \qquad\mbox{and}\qquad
f=e^x \quad\mbox{or}\quad (f,h)=(1,0) \quad\mbox{otherwise}
\]
seems to be more complicated than $h=0$. But it is the gauge that
straight leads to exhaustive classification, and formulation of
results having a simplest form (Cases~8--13 of Table~1).
\end{example}

\begin{note}\label{NoteOnEarlyApplicationofEquivTransformations}
We tested different ways of classification.
Implemented `experiments' result in the following conclusion.
Application of equivalence transformations is more effective on earlier stages of classification.
An expedient way is to gauge arbitrary elements as much as possible on preparatory classification stage
and additionally to gauge arbitrary elements under classification every time
when sufficient conditions to do it arises.
The worst choice is to refuse to use gauges completely.
\end{note}

After analyzing the classification adduced in Table~1, we can formulate the following theorem
which is similar to a one from~\cite{Popovych&Ivanova2004NVCDCEs}.

\begin{theorem}\label{TheoremOnRDfghEqsWith4DimLieInvAlgs}
If an equation of form~(\ref{eqRDfghPower}) is invariant with respect to a Lie algebra
of dimension not less than four then it can be reduced by means of point transformations to
a one with constant values of the parameters $f$, $g$ and $h$.
If $m\not=1,n+1$, the similar statement is also true for three-dimensional Lie invariance algebras.
\end{theorem}

\section{Additional equivalence transformations}\label{SectionOnAddEqiuvTrans}

Let a class of differential equations be classified with respect to its equivalence group $G^{\sim}$
and projection of $G^{\sim}$ on the space of the equation variables be narrower than the whole pseudo-group of point transformations in this space.
Then there can exist point transformations between $G^{\sim}$-inequivalent cases of symmetry extension,
which are called \emph{additional equivalence transformations}.
Knowledge of them is important since it simplifies further application of group classification results.

The first nontrivial and interesting example of such transformations was given in~\cite{kurdyumov1989} (see
also~\cite[Chapter~10]{Ibragimov1994V1}).
Under classification of `constant coefficient' nonlinear diffusion--reaction equations~\eqref{EqNCCDR} in~\cite{Dorodnitsyn1979,Dorodnitsyn1982},
equations of the form
\[
u_t=(u^{-4/3}u_x)_x+bu^{-1/3},
\]
where $b$ is an arbitrary constant, arose as cases of  symmetry extension,
which are inequivalent with respect to the corresponding equivalence group for different values of~$b$.
It was shown in ~\cite{kurdyumov1989} that the transformation
\[
t'=t,
\quad
x'=\left \{\begin{array}{ll} e^{-2\sqrt{b/3}x},&b>0 \\ \tan (\sqrt{-b/3}x),&b<0 \end{array} \right.,
\quad
u'=\left \{\begin{array}{ll} e^{3\sqrt{b/3}x}u,&b>0 \\ \cos^3 (\sqrt{-b/3}x) u,&b<0 \end{array} \right.
\]
maps any such equation with the equation
\[
u'_{t'}=(u^{\prime -4/3}u'_{x'})^{}_{x'}
\]
which is of the same form with $b'=0$. Therefore, it is an
additional equivalence transformation in class~\eqref{EqNCCDR}. It
is an additional equivalence transformation in
classes~\eqref{eqRDfghPower} and~\eqref{eqRDfghPowerG1} and, at
the same time, a member of the equivalence group~$G^{\sim}_{1,
m=n+1}$ of class~\eqref{eqRDfghPowerG1MN1}. That is why the case
$b\not=0$ does not arise under our classification. Let us note
that application of generalized, extended and conditional
equivalence groups simultaneously with the usual one results in
implicit but effective usage of additional equivalences.

Another additional equivalence transformation in class~\eqref{EqNCCDR} \cite[Chapter~10]{Ibragimov1994V1}
\begin{equation}\label{EqAddEquivTransOfEqRDfghPower}
t'=\frac 1{\varepsilon n}e^{\varepsilon n t},\quad
x'=x,\quad
u'=e^{-\varepsilon t}u
\end{equation}
links the equations
\begin{equation}\label{EqNCCDRConnectedWithAddEquivTrans}
u_t=(u^{n}u_x)_x+\varepsilon u
\quad\mbox{and}\quad
u'_{t'}=(u^{\prime n}u'_{x'})^{}_{x'}.
\end{equation}
This transformation belongs to no equivalence group found in Section~\ref{SectionOnEquivTransOfEqRDfghPower}.
It can be obviously extended to a wider subclass of~\eqref{eqRDfghPower}.
Namely, it connects the equations
\[
f(x)u_t=(g(x)u^{n}u_x)_x+\varepsilon f(x)u
\quad\mbox{and}\quad
f(x')u'_{t'}=(g(x')u^{\prime n}u'_{x'})^{}_{x'}
\]
and, therefore, reduces Cases~4--7 of Table~1 to the set of cases
`$m=n+1$ or $h=0$', i.e. it is an additional equivalence
transformation in class~\eqref{eqRDfghPower}. Any other additional
equivalence transformations is a composition of transformations of
form~\eqref{EqAddEquivTransOfEqRDfghPower} and transformations
from~$G^{\sim}_{1, m=n+1}$. (It~is proved in the next section in
the framework of admissible transformations.) For example, if
$n=-\frac 43$ equations~\eqref{EqNCCDRConnectedWithAddEquivTrans}
are also connected by the mapping
\[
t'=\frac 1{\varepsilon n}e^{\varepsilon n t},\quad
x'=\frac{\delta}x,\quad u'=e^{-\varepsilon
t}x^3u,\quad \quad \delta =\pm 1.
\]

Seeming scarcity of additional equivalence transformations in class~\eqref{eqRDfghPower},
e.g. in comparison with the class $f(x)u_t=[g(x)A(u)u_x]_x+h(x)B(u)u_x$ admitting a large number of additional equivalence transformations
\cite{Ivanova&Sophocleous2006,Popovych&Ivanova2004NVCDCEs}, is connected with usage of conditional equivalence groups under classification.
As it is demonstrated in the next section, the transformations of form~\eqref{EqAddEquivTransOfEqRDfghPower}
can be also included in the framework of conditional equivalence.

As a result, the following statement is derived.

\begin{theorem}\label{TheoremOnClassificationOfRDfghEqsWRTPointTrans}
Up to point transformations, a complete list of extensions of the maximal Lie invariance group of equations from class~\eqref{eqRDfghPower}
is exhausted by Cases~1--3 and~8--13 of Table~1.
\end{theorem}

\section{Classification of form-preserving (admissible) transformations}\label{SectionClassificationOfAdmTrans}

Due to special structure of equations from
class~\eqref{eqRDfghPower}, the following problem can be solved
completely. To describe all point transformations each of which
connects a pair of equations from class~\eqref{eqRDfghPower}. Such
transformations are called
\emph{form-preserving}~\cite{Kingston&Sophocleous1998} or
\emph{admissible}~\cite{Popovych&Eshraghi2005}
\emph{transformations}. See~\cite{Popovych&Eshraghi2005} for
stronger definitions. They can be naturally interpreted in terms
of the category theory~\cite{Prokhorova2005}. 
Let us note that there exists an infinitesimal equivalent of this notion~\cite{Borovskikh2004}.

Since class~\eqref{eqRDfghPower} can be gauged with transformations from its usual equivalence group~$G^{\sim}$,
it is enough for one to solve the similar problem in class~\eqref{eqRDfghPowerG1}.
To do it, we consider a pair of equations from the class under consideration, i.e. \eqref{eqRDfghPowerG1} and
\begin{equation}\label{eqRDfghPowerG1a}
\tilde f(\tilde x)\tilde u_{\tilde t}=
(\tilde u^{\tilde n}\tilde u_{\tilde x})_{\tilde x}+\tilde h(\tilde x)\tilde u^{\tilde m}
\end{equation}
and assume that these equations are connected via a point transformation of the general form
\[
\tilde t=T(t,x,u), \quad
\tilde x=X(t,x,u), \quad
\tilde u=U(t,x,u).
\]
We have to derive determining equations in functions~$T$, $X$ and $U$ and to solve them
depending on values of arbitrary elements in~\eqref{eqRDfghPowerG1} and~\eqref{eqRDfghPowerG1a}.

After substitution of expressions for the tilde-variables into~\eqref{eqRDfghPowerG1a}, we obtain an equation
in the tildeless variables. It should be an identity on the manifold determined by~\eqref{eqRDfghPowerG1}
in the second-order jet space over the variables $(t,x\,|\,u)$.
Splitting of this identity with respect to the unconstrained variables in a true way
results at first in the equations
\[
T_u=T_x=X_u=U_{uu}=0
\]
that agrees with results on more general classes of
evolution equations~\cite{Kingston&Sophocleous1998,Popovych&Ivanova2004NVCDCEs,Prokhorova2005}.
We split further, taking into account the above equations and also splitting with respect to~$u$ partially.
As a results, the following relations are deduced:
\begin{gather}
\tilde n=n,\qquad \tilde m=m \quad\mbox{or}\quad (m,\tilde m)\in\{(1,n+1),(n+1,1)\},\nonumber
\\[1ex]
T=T(t),\quad X=X(x),\quad U=V(t,x)u,\quad T_tX_xV\not=0\nonumber
\\[1ex]
2(n+1)X_xV_x=X_{xx}V,\nonumber
\\[1ex]
V^n=\frac{\tilde f}f \frac{X_x{}^2}{T_t},\nonumber
\\[1ex]
\frac{\tilde f}f \frac{V}{T_t}hu^m+\tilde f\frac{V_t}{T_t}u=
\frac1{X_x}\left(V^n\frac{V_x}{X_x}\right)_xu^{n+1}+\tilde hV^{\tilde m}u^{\tilde m}.
\label{EqRDfghPowerForClassificationOfAdmTrans}
\end{gather}

Splitting of equation~\eqref{EqRDfghPowerForClassificationOfAdmTrans} with respect to~$u$
and subsequent integration of the determining equations depends on values of~$m$ and $\tilde m$ appreciably.
Consider different cases separately.

If $\tilde m=m\not=1$,
equation~\eqref{EqRDfghPowerForClassificationOfAdmTrans} implies
$V_t=0$ hence $T_{tt}=0$, i.e. $V=V(x)$ and $T= \delta_1t+\delta_2$. The other conditions are
\begin{gather*}
\tilde m=m\not=1,n+1\colon\quad
\frac{\tilde f}f =\delta_1\frac{V^n}{X_x}, \quad
\frac{\tilde h}h =\frac{V^{n+1-m}}{X_x{}^2},
\\[.5ex]
\qquad n=-1\colon\quad X_{xx}=0, \quad (\ln V)_{xx}=0,
\\[.5ex]
\qquad n\not=-1\colon\quad \bigl(|X_x|^{-1/2}\bigr)_{xx}=0, \quad V^{2n+2}=\delta_0X_x;
\\[1.5ex]
\tilde m=m=n+1\colon\quad
\frac{\tilde f}f =\delta_1\frac{V^n}{X_x}, \quad
\tilde h=\frac1{X_x{}^2}h-\frac1{X_xV^{n+1}}\left(V^n\frac{V_x}{X_x}\right)_x,\quad
V^{2n+2}=\delta_0X_x.
\end{gather*}
Therefore,  in the case $m\not=1,n+1$ (the case $m=n+1$) any admissible transformation is determined by
a transformation from the equivalence group~$G^{\sim}_1$ (the conditional equivalence group~$G^{\sim}_{1, m=n+1}$).

If $\tilde m=m=1$, the function~$V$ is conveniently presented as $V=|T_t|^{-1/n}\psi(x)$.
The value $n=-1$ is again singular:
\begin{gather*}
n=-1\colon\quad X_{xx}=0, \ (\ln \psi)_{xx}=0
\quad \Longrightarrow \quad\!
X=\delta_3x+\delta_4,\ \psi=\delta_5e^{\delta_6x},
\\[.5ex]
n\not=-1\colon\quad \bigl(|X_x|^{-1/2}\bigr)_{xx}=0, \ 2(n+1)\frac{\psi_x}{\psi}=\frac{X_{xx}}{X_x}
\quad \Longrightarrow \quad\!
X=\frac{\delta_3x+\delta_4}{\delta_5x+\delta_6},\ \psi=\delta_7|X_x|^{\frac1{2n+2}},
\end{gather*}
where $\delta_3\delta_6-\delta_4\delta_5$ has a fixed value which
can be assumed equal to $\pm1$. The arbitrary elements~$f$ and~$h$
are transformed by the formulas
\[
\frac{\tilde f}f =\frac{\psi^n}{X_x{}^2}\sign T_t,
\quad
\frac{\tilde h}{\tilde f}=\frac1{T_t}\frac{h}{f}+\frac1n\left(\frac1{T_t}\right)_t,
\quad\mbox{hence}\quad
\biggl(\frac{\tilde h}{\tilde f}\biggr)_x=\frac1{T_t}\biggl(\frac{h}{f}\biggr)_x.
\]
In view of the latter condition, $(\tilde h/\tilde f)_x\not=0$ and $T_{tt}=0$ if $(h/f)_x\not=0$.
Therefore, any admissible transformation in this case corresponds to
a transformation from the equivalence group~$G^{\sim}_1$.

Different situation is in case of $(h/f)_x=0$. Then $(\tilde
h/\tilde f)_x=0$, i.e. the condition $(h/f)_x=0$ is invariant with
respect to~$G^{\sim}_1$. Let the constants $h/f$ and $\tilde
h/\tilde f$ be denoted by~$\alpha$ and~$\tilde\alpha$,
respectively. Then we get the following equation for the
function~$T(t)$:
\[
\left(\frac1{T_t}\right)_t=-n\alpha\frac1{T_t}+n\tilde\alpha.
\]
We integrate these equation and present the general solution in such form that
continuous dependence of it on the parameters~ $\alpha$ and~$\tilde\alpha$ is obvious:
\begin{gather*}
\alpha\tilde\alpha\not=0\colon \quad
\frac{e^{n\tilde\alpha T}-1}{n\tilde\alpha}=\delta_1\frac{e^{n\alpha t}-1}{n\alpha}+\delta_2,
\qquad
\alpha=0,\ \tilde\alpha\not=0\colon \quad \frac{e^{n\tilde\alpha T}-1}{n\tilde\alpha}=\delta_1t+\delta_2,
\\[.5ex]
\alpha\not=0,\ \tilde\alpha=0\colon \quad T=\delta_1\frac{e^{n\alpha t}-1}{n\alpha}+\delta_2,
\qquad
\alpha=\tilde\alpha=0\colon \quad T=\delta_1t+\delta_2.
\end{gather*}

The above-described transformations form the equivalence group $G^{\sim}_{1,m=1,(h/f)_x=0}$ of
the subclass of class~\eqref{eqRDfghPower}, which is separated with
the conditions $g=1$, $m=1$, $(h/f)_x=0$. It is a generalized equivalence group
even through $n$ is fixed since it contains transformations with respect to~$t$,
which depends on the (constant) ratio of the arbitrary elements~$h$ and~$f$.
Unlike $G^{\sim}_{1,m=n+1}$, we do not use $G^{\sim}_{1,m=1,(h/f)_x=0}$ under group classification
of class~\eqref{eqRDfghPowerG1}
since application of this conditional equivalence group does not have crucial influence on classification
and the corresponding conditions on arbitrary elements are less obvious.
At the same time, transformations from $G^{\sim}_{1,m=1,(h/f)_x=0}$ play the role of
additional equivalence transformations after completing the classification (see the previous section).

If $h\tilde h\not=0$, $m=1$ and $\tilde m=n+1$ then the
determining equations yield
\[
\frac hf=\frac1n\frac{T_{tt}}{T_t}, \quad\mbox{i.e}\quad \left(\frac{h}{f}\right)_x=0,
\qquad
\tilde h=-\frac1{X_xV^{n+1}}\left(V^n\frac{V_x}{X_x}\right)_x.
\]
Hence both equations~\eqref{eqRDfghPowerG1}
and~\eqref{eqRDfghPowerG1a} are mapped to the subclass `$h=0$' of
the class under consideration. These mappings are realized by the
transformations from the corresponding conditional equivalence
groups. We can assume that their images coincides. Therefore, the
admissible transformation is the composition of a
transformation~$\mathcal T_1$ of~\eqref{eqRDfghPowerG1} to the
equation $f(x)u_t=(u^nu_x)_x$ and a transformation~$\mathcal T_2$ of
the equation $f(x)u_t=(u^nu_x)_x$ to~\eqref{eqRDfghPowerG1a}. The
transformation~$\mathcal T_1$ belongs to
$G^{\sim}_{1,m=1,(h/f)_x=0}$ and the transformation~$\mathcal T_2$
belongs to $G^{\sim}_{1,m=n+1}$, and we can put $X=x$, $\psi=1$
in~$\mathcal T_1$.

The case $h\tilde h\not=0$, $m=n+1$ and $\tilde m=1$ is considered in similar way.

All the possible cases are exhausted.

Let us summarize investigation of the admissible transformations
in class~\eqref{eqRDfghPowerG1} in  the following statement.

\begin{theorem}\label{TheoremOnRDfghSetOfAdmTrans}
Let the equations $f(x)u_t=(u^nu_x)_x + h(x)u^m$ and\,
$\tilde f(x)u_{t}=(u^{\tilde n}u_x)_x+\tilde h(x)u^{\tilde m}$
be connected via a point transformation in the variables~$t$, $x$ and~$u$.
Then
\[
\tilde n=n \quad\mbox{and}\quad \quad\mbox{either}\quad
\tilde m=m \quad\mbox{or}\quad (m,\tilde m)=(1,n+1) \quad\mbox{or}\quad (m,\tilde m)=(n+1,1).
\]
The transformation is determined by a transformation from

\medskip

a) $G^{\sim}_1$ if either $m\not=1,n+1$ or $m=1$, $(h/f)_x\not=0$;

b) $G^{\sim}_{1,m=n+1}$ if $m=\tilde m=n+1$;

c) $G^{\sim}_{1,m=1,(h/f)_x=0}$ if $m=\tilde m=1$, $(h/f)_x=0$; then also $(\tilde h/\tilde f)_x=0$.

\medskip

If $m=1$ and $\tilde m=n+1$ then $(h/f)_x=0$ and the transformation is composition of
two transformations from $G^{\sim}_{1,m=1,(h/f)_x=0}$ and $G^{\sim}_{1,m=n+1}$ with
the intermediate equation having $h=0$.

The case  $m=n+1$ and $\tilde m=1$ is similar.
\end{theorem}

\begin{corollary}\label{CorollaryOnRDfghSetOfAdmTrans}
Class~\eqref{eqRDfghPowerG1} with $n\not=0$ can be presented as the union of normalized subclasses separated by the conditions
\[
h\not=0, \ m\not=1,n+1;\qquad
m=1,\ (h/f)_x\not=0;\qquad
m=1, \ (h/f)_x=0;\qquad
m=n+1.
\]
Only two latter subclasses have a non-empty intersection, and the intersection being the normalized subclass `$h=0$'.
\end{corollary}

Roughly speaking, the class of differential equations is called normalized if any
admissible transformation in this class belongs to its equivalence group. See~\cite{Popovych&Eshraghi2005} for strong definitions.

\section{Basic definitions and statements on conservation laws}\label{SectionOnBasicDefsOfCLs}

In this section we give basic definitions and statements on
conservation laws, following the well-known textbook of Olver~\cite{Olver1986} in general outlines. 
Then we formulate the notion of equivalence of
conservation laws with respect to equivalence groups, which was
first introduced in~\cite{Popovych&Ivanova2004ConsLawsLanl}. This
notion is a base for modification of the direct method of
construction of conservation laws, which is applied in
Section~\ref{SectionOnRDfghConsLaws} for exhaustive classification of
local conservation laws of equations~(\ref{eqRDfghPower}).

Let~$\mathcal{L}$ be a system~$L(x,u_{(\rho)})=0$ of $l$
differential equations $L^1=0$, \ldots, $L^l=0$ for $m$ unknown
functions $u=(u^1,\ldots,u^m)$ of $n$ independent variables
$x=(x_1,\ldots,x_n).$ Here $u_{(\rho)}$ denotes the set of all the
derivatives of the functions $u$ with respect to $x$ of order no
greater than~$\rho$, including $u$ as the derivatives of the zero
order. Let $\mathcal{L}_{(k)}$ denote the set of all algebraically
independent differential consequences that have, as differential
equations, orders no greater than $k$. We
identify~$\mathcal{L}_{(k)}$ with the manifold determined
by~$\mathcal{L}_{(k)}$ in the jet space~$J^{(k)}$.

\begin{definition}\label{def.conservation.law}
A {\em conserved vector} of the system~$\mathcal{L}$ is an
$n$-tuple $F=(F^1(x,u_{(r)}),\ldots,F^n(x,u_{(r)}))$ for which the
divergence ${\rm Div}\,F:=D_iF^i$ vanishes for all solutions
of~$\mathcal{L}$ (i.e. ${\rm Div}F\bigl|_\mathcal{L}=0$).
\end{definition}

In Definition~\ref{def.conservation.law} and below $D_i=D_{x_i}$
denotes the operator of total differentiation with respect to the
variable~$x_i$, i.e. $D_i=\p_{x_i}+u^a_{\alpha,i}\p_{u^a_\alpha}$,
where $u^a_\alpha$ and $u^a_{\alpha,i}$ stand for the variables in
jet spaces, which correspond to derivatives $\p^{|\alpha|}u^a/\p
x_1^{\alpha_1}\ldots\p x_n^{\alpha_n}$ and $\p u^a_\alpha/\p x_i$,
$\alpha=(\alpha_1,\ldots,\alpha_n)$,
$\alpha_i\in\mathbb{N}\cup\{0\}$,
$|\alpha|{:}=\alpha_1+\cdots+\alpha_n$. We use the summation
convention for repeated indices and assume any function as its
zero-order derivative. The notation~$V\bigl|_\mathcal{L}$ means
that values of $V$ are considered only on solutions of the
system~$\mathcal{L}$.

\begin{definition}
A conserved vector $F$ is called {\em trivial} if $F^i=\hat
F^i+\check F^i,$ $i=\overline{1,n},$ where $\hat F^i$ and $\check
F^i$ are, likewise $F^i$, functions of $x$ and derivatives of $u$
(i.e. differential functions), $\hat F^i$ vanish on the solutions
of~$\mathcal L$ and the $n$-tuple $\check F=(\check
F^1,\ldots,\check F^n)$ is a null divergence (i.e. its divergence
vanishes identically).
\end{definition}

The triviality concerning the vanishing conserved vectors on
solutions of the system can be easily eliminated by confining on
the manifold of the system, taking into account all its necessary
differential consequences. A characterization of all null
divergences is given by the following lemma (see
e.g.~\cite{Olver1986}).

\begin{lemma}\label{lemma.null.divergence}
The $n$-tuple $F=(F^1,\ldots,F^n)$, $n\ge2$, is a null divergence
($\mathop{\rm Div}\nolimits F\equiv0$) iff there exist smooth
functions $v^{ij}$ ($i,j=\overline{1,n}$) of $x$ and derivatives
of $u$, such that $v^{ij}=-v^{ji}$ and $F^i=D_jv^{ij}$.
\end{lemma}

The functions $v^{ij}$ are called {\em potentials} corresponding
to the null divergence~$F$. If $n=1$ any null divergence is
constant.

\begin{definition}\label{DefinitionOfConsVectorEquivalence}
Two conserved vectors $F$ and $F'$ are called {\em equivalent} if
the vector-function $F'-F$ is a trivial conserved vector.
\end{definition}

The above definitions of triviality and equivalence of conserved
vectors are natural in view of the usual `empiric' definition of
conservation laws of a system of differential equations as
divergences of its conserved vectors, i.e. divergence expressions
which vanish for all solutions of this system. For example,
equivalent conserved vectors correspond to the same conservation
law. It allows us to formulate the definition of conservation law
in a rigorous style (see e.g.~\cite{Zharinov1986}). Namely, for
any system~$\mathcal{L}$ of differential equations the
set~$\CV(\mathcal{L})$ of conserved vectors of its conservation
laws is a linear space, and the subset~$\CV_0(\mathcal{L})$ of
trivial conserved vectors is a linear subspace
in~$\CV(\mathcal{L})$. The factor
space~$\CL(\mathcal{L})=\CV(\mathcal{L})/\CV_0(\mathcal{L})$
coincides with the set of equivalence classes
of~$\CV(\mathcal{L})$ with respect to the equivalence relation
adduced in Definition~\ref{DefinitionOfConsVectorEquivalence}.

\begin{definition}\label{DefinitionOfConsLaws}
The elements of~$\CL(\mathcal{L})$ are called {\em conservation
laws} of the system~$\mathcal{L}$, and the whole factor
space~$\CL(\mathcal{L})$ is called as {\em the space of
conservation laws} of~$\mathcal{L}$.
\end{definition}

That is why we assume description of the set of conservation laws
as finding~$\CL(\mathcal{L})$ that is equivalent to construction
of either a basis if $\dim \CL(\mathcal{L})<\infty$ or a system of
generatrices in the infinite dimensional case. The elements
of~$\CV(\mathcal{L})$ which belong to the same equivalence class
giving a conservation law~${\cal F}$ are considered all as
conserved vectors of this conservation law, and we will
additionally identify elements from~$\CL(\mathcal{L})$ with their
representatives in~$\CV(\mathcal{L})$. For $F\in\CV(\mathcal{L})$
and ${\cal F}\in\CL(\mathcal{L})$ the notation~$F\in {\cal F}$
will denote that $F$ is a conserved vector corresponding to the
conservation law~${\cal F}$. In contrast to the order $r_F$ of a
conserved vector~$F$ as the maximal order of derivatives
explicitly appearing in~$F$, the {\em order of the conservation
law}~$\cal F$ is called $\min\{r_F\,|\,F\in{\cal F}\}$. Under
linear dependence of conservation laws we understand linear
dependence of them as elements of~$\CL(\mathcal{L})$. Therefore,
in the framework of ``representative'' approach conservation laws
of a system~$\mathcal{L}$ are considered {\em linearly dependent}
if there exists linear combination of their representatives, which
is a trivial conserved vector.

Let the system~$\cal L$ be totally nondegenerate~\cite{Olver1986}.
Then application of the Hadamard lemma to the definition of
conserved vector and integrating by parts imply that divergence of
any conserved vector of~$\mathcal L$ can be always presented, up
to the equivalence relation of conserved vectors, as a linear
combination of left side of independent equations from $\mathcal
L$ with coefficients~$\lambda^\mu$ being functions on a suitable
jet space~$J^{(k)}$:
\begin{equation}\label{CharFormOfConsLaw}
\mathop{\rm Div}\nolimits F=\lambda^\mu L^\mu.
\end{equation}
Here the order~$k$ is determined by~$\mathcal L$ and the allowable
order of conservation laws, $\mu=\overline{1,l}$.

\begin{definition}\label{DefCharForm}
Formula~\eqref{CharFormOfConsLaw} and the $l$-tuple
$\lambda=(\lambda^1,\ldots,\lambda^l)$ are called the {\it
characteristic form} and the {\it characteristic} of the
conservation law~$\mathop{\rm Div}\nolimits F=0$ correspondingly.
\end{definition}

The characteristic~$\lambda$ is {\em trivial} if it vanishes for
all solutions of $\cal L$. Since $\cal L$ is nondegenerate, the
characteristics~$\lambda$ and~$\tilde\lambda$
satisfy~\eqref{CharFormOfConsLaw} for the same~$F$ and, therefore,
are called {\em equivalent} iff $\lambda-\tilde\lambda$ is a
trivial characteristic. Similarly to conserved vectors, the
set~$\Ch(\mathcal{L})$ of characteristics corresponding to
conservation laws of the system~$\cal L$ is a linear space, and
the subset~$\Ch_0(\mathcal{L})$ of trivial characteristics is a
linear subspace in~$\Ch(\mathcal{L})$. The factor space~$\Ch_{\rm
f}(\mathcal{L})=\Ch(\mathcal{L})/\Ch_0(\mathcal{L})$ coincides
with the set of equivalence classes of~$\Ch(\mathcal{L})$ with
respect to the above characteristic equivalence relation.

We can essentially simplify and order classification of
conservation laws, taking into account additionally symmetry
transformations of a system or equivalence transformations of a
whole class of systems. Such problem is similar to one of group
classification of differential equations.

\begin{proposition}
Any point transformation~$g$ maps a class of equations in the
conserved form into itself. More exactly, the transformation~$g$:
$\tilde x=x_g(x,u)$, $\tilde u=u_g(x,u)$ prolonged to the jet
space~$J^{(r)}$ transforms the equation $D_iF^i=0$ to the equation
$D_iF^i_g=0$. The transformed conserved vector~$F_g$ is determined
by the formula
\begin{equation}\label{eq.tr.var.cons.law}
F_g^i(\tilde x,\tilde u_{(r)})=\frac{D_{x_j}\tilde x_i}{|D_x\tilde
x|}\,F^j(x,u_{(r)}), \quad\mbox{i.e.}\quad F_g(\tilde x,\tilde
u_{(r)})=\frac{1}{|D_x\tilde x|}(D_x\tilde x)F(x,u_{(r)})
\end{equation}
in the matrix notions. Here $|D_x\tilde x|$ is the determinant of
the matrix $D_x\tilde x=(D_{x_j}\tilde x_i)$.
\end{proposition}

\begin{note}
In the case of one dependent variable ($m=1$) $g$ can be a contact
transformation: $\tilde x=x_g(x,u_{(1)})$, $\tilde
u_{(1)}=u_{g(1)}(x,u_{(1)})$. Similar notes are also true for the
statements below.
\end{note}

\begin{definition}
Let $G$ be a symmetry group of the system~$\mathcal{L}$. Two
conservation laws with the conserved vectors $F$ and $F'$ are
called {\em $G$-equivalent} if there exists a transformation $g\in
G$ such that the conserved vectors $F_g$ and $F'$ are equivalent
in the sense of
Definition~\ref{DefinitionOfConsVectorEquivalence}.
\end{definition}

Any transformation $g\in G$ induces a linear one-to-one mapping
$g_*$ in~$\CV(\mathcal{L})$, transforms trivial conserved vectors
only to trivial ones (i.e. $\CV_0(\mathcal{L})$ is invariant with
respect to~$g_*$) and therefore induces a linear one-to-one
mapping $g_{\rm f}$ in~$\CL(\mathcal{L})$. It is obvious that
$g_{\rm f}$ preserves linear (in)dependence of elements
in~$\CL(\mathcal{L})$ and maps a basis (a set of generatrices)
of~$\CL(\mathcal{L})$ in a basis (a set of generatrices) of the
same space. In such way we can consider the $G$-equivalence
relation of conservation laws as well-determined
on~$\CL(\mathcal{L})$ and use it to classify conservation laws.

\begin{proposition}
If the system~$\mathcal{L}$ admits a one-parameter group of
transformations then the infinitesimal generator
$X=\xi^i\p_i+\eta^a\p_{u^a}$ of this group can be used for
construction of new conservation laws from known ones. Namely,
differentiating equation~(\ref{eq.tr.var.cons.law}) with respect
to the parameter $\varepsilon$ and taking the value
$\varepsilon=0$, we obtain the new conserved vector
\begin{equation}\label{eq.inf.tr.var.cons.law}
\widetilde F^i=-X_{(r)}F^i+(D_j\xi^i)F^j-(D_j\xi^j)F^i.
\end{equation}
Here $X_{(r)}$ denotes the $r$-th
prolongation~\cite{Olver1986,Ovsiannikov1982} of the operator $X$.
\end{proposition}

\begin{note}Formula~\eqref{eq.inf.tr.var.cons.law} can be directly extended to generalized symmetry operators
(see, for example,~\cite{Anco&Bluman2002c}). A
similar statement for generalized symmetry operators in
evolutionary form ($\xi^i=0$) was known
earlier~\cite{Ibragimov1985,Olver1986}. It was used
in~\cite{Khamitova1982} to introduce a notion of basis of
conservation laws as a set which generates a whole set of
conservation laws with action of generalized symmetry operators
and operation of linear combination.
\end{note}

\begin{proposition}\label{PropositionOnInducedMapping}
Any point transformation $g$ between systems~$\mathcal{L}$
and~$\tilde{\mathcal{L}}$ induces a linear one-to-one mapping
$g_*$ from~$\CV(\mathcal{L})$ into~$\CV(\tilde{\mathcal{L}})$,
which maps $\CV_0(\mathcal{L})$ into~$\CV_0(\tilde{\mathcal{L}})$
and generates a linear one-to-one mapping $g_{\rm f}$
from~$\CL(\mathcal{L})$ into~$\CL(\tilde{\mathcal{L}})$.
\end{proposition}

\begin{corollary}\label{CorollaryOnInducedMappingOfChar}
Any point transformation $g$ between systems~$\mathcal{L}$
and~$\tilde{\mathcal{L}}$ induces a linear one-to-one mapping
$\hat g_{\rm f}$ from~$\Ch_{\rm f}(\mathcal{L})$ into~$\Ch_{\rm
f}(\tilde{\mathcal{L}})$.
\end{corollary}
It is possible to obtain an explicit formula for correspondence
between characteristics of~$\mathcal{L}$
and~$\tilde{\mathcal{L}}$. Let
$\tilde{\mathcal{L}}^\mu=\Lambda^{\mu\nu}\mathcal{L}^\nu$, where
$\Lambda^{\mu\nu}=\Lambda^{\mu\nu\alpha}D^\alpha$,
$\Lambda^{\mu\nu\alpha}$ are differential functions,
$\alpha=(\alpha_1,\ldots,\alpha_n)$ runs the multi-indices set
($\alpha_i\!\in\!\mathbb{N}\cup\{0\}$), $\mu,\nu=\overline{1,l}$.
Then
\[\lambda^\mu={\Lambda^{\nu\mu}}^*(|D_x\tilde x|\tilde\lambda^\nu).\]
Here ${\Lambda^{\nu\mu}}^*=(-D)^\alpha\cdot\Lambda^{\mu\nu\alpha}$
is the adjoint to the operator~$\Lambda^{\nu\mu}$. For a number of
cases, e.g. if~$\mathcal{L}$ and~$\tilde{\mathcal{L}}$ are single
partial differential equations ($l=1$), the
operators~$\Lambda^{\mu\nu}$ are simply differential functions
(i.e. $\Lambda^{\mu\nu\alpha}=0$ for $|\alpha|>0$) and, therefore,
${\Lambda^{\nu\mu}}^*=\Lambda^{\mu\nu}$.

Consider the class~$\mathcal{L}|_{\cal S}$ of
systems~$\mathcal{L}_\theta$:
$L(x,u_{(\rho)},\theta(x,u_{(\rho)}))=0$ parameterized with the
parameter-functions~$\theta=\theta(x,u_{(\rho)}).$ Here $L$ is a
tuple of fixed functions of $x,$ $u_{(\rho)}$ and $\theta.$
$\theta$~denotes the tuple of arbitrary (parametric) functions
$\theta(x,u_{(\rho)})=(\theta^1(x,u_{(\rho)}),\ldots,\theta^k(x,u_{(\rho)}))$
running the set~${\cal S}$ of solutions of the
system~$S(x,u_{(\rho)},\theta_{(q)}(x,u_{(\rho)}))=0$. This system
consists of differential equations on $\theta$, where $x$ and
$u_{(\rho)}$ play the role of independent variables and
$\theta_{(q)}$ stands for the set of all the partial derivatives
of $\theta$ of order no greater than $q$. In what follows we call
the functions $\theta$ arbitrary elements. Denote the point
transformations group preserving the form of the systems
from~$\mathcal{L}|_{\cal S}$ as $G^{\Equiv}=G^{\Equiv}(L,S).$

Consider the set~$P=P(L,S)$ of all pairs each of which consists of
a system $\mathcal{L}_\theta$ from~$\mathcal{L}|_{\cal S}$ and a
conservation law~${\cal F}$ of this system. In view of
Proposition~\ref{PropositionOnInducedMapping}, action of
transformations from~$G^{\Equiv}$ on $\mathcal{L}|_{\cal S}$ and
$\{\CV(\mathcal{L}_{\theta})\,|\,\theta\in{\cal S}\}$ together
with the pure equivalence relation of conserved vectors naturally
generates an equivalence relation on~$P$.

\begin{definition}
Let $\theta,\theta'\in{\cal S}$, ${\cal
F}\in\CL(\mathcal{L}_\theta)$, ${\cal
F}'\in\CL(\mathcal{L}_{\theta'})$, $F\in{\cal F}$, $F'\in{\cal
F'}$. The pairs~$(\mathcal{L}_\theta,{\cal F})$
and~$(\mathcal{L}_{\theta'},{\cal F'})$ are called {\em
$G^{\Equiv}$-equivalent} if there exists a transformation $g\in
G^{\Equiv}$ which transforms the system~$\mathcal{L}_\theta$ to
the system~$\mathcal{L}_{\theta'}$ and such that the conserved
vectors $F_g$ and $F'$ are equivalent in the sense of
Definition~\ref{DefinitionOfConsVectorEquivalence}.
\end{definition}

Classification of conservation laws with respect to~$G^{\Equiv}$
will be understood as classification in~$P$ with respect to the
above equivalence relation. This problem can be investigated in
the way that is similar to group classification in classes of
systems of differential equations, especially it is formulated in
terms of characteristics. Namely, we construct firstly the
conservation laws that are defined for all values of the arbitrary
elements. (The corresponding conserved vectors may depend on the
arbitrary elements.) Then we classify, with respect to the
equivalence group, arbitrary elements for each of that the system
admits additional conservation laws.

In an analogues way we also can introduce equivalence relations
on~$P$, which are generated by either generalizations of usual
equivalence groups or all admissible point or contact
transformations (called also form-preserving
ones~\cite{Kingston&Sophocleous1998}) in pairs of equations
from~$\mathcal{L}|_{\cal S}$.

\begin{note}
It can be easy shown that all the above equivalences are indeed
equivalence relations, i.e., they have the usual reflexive,
symmetric and transitive properties.
\end{note}

\section{Conservation laws}\label{SectionOnRDfghConsLaws}

We search (local) conservation laws of equations from class
(\ref{eqRDfghPower}), applying the modification of the most direct
method, which was proposed in~\cite{Popovych&Ivanova2004ConsLawsLanl}
and preliminaries of which are described in the previous section.
It is based on using the definition of conservation laws,
the notion of equivalence of conservation laws with respect to a
transformation group and classification up to the equivalence
group of a class of differential equations.
In view of results of the previous section,
 it is sufficient for exhaustive investigation if
we classify conservation laws of equations only from
class~(\ref{eqRDfghPowerG1}). Let us note that other kinds of the
direct methods, which are based e.g. on the characteristic form of
conservation laws~\cite{Anco&Bluman2002a,Anco&Bluman2002b} could
be also applied.

There are two independent variables~$t$ and~$x$ in equations under
consideration, which play a part of the time and space variables
correspondingly. Therefore, the general form of constructed
conservation laws will be
\begin{equation}\label{EqGen2DimConsLaw}
D_tF(t,x,u_{(r)})+D_xG(t,x,u_{(r)})=0,
\end{equation}
where $D_t$ and $D_x$ are the operators of total differentiation
with respect to $t$ and $x$. The components $F$ and $G$ of the
conserved vector~$(F,G)$ are called the {\em density} and the {\em
flux} of the conservation law. Two conserved vectors $(F,G)$ and
$(F',G')$ are equivalent if there exist such functions~$\hat F$,
$\hat G$ and~$H$ of~$t$, $x$ and derivatives of~$u$ that $\hat F$
and $\hat G$ vanish for all solutions of~$\mathcal{L}$~and
$F'=F+\hat F+D_xH$, $G'=G+\hat G-D_tH$.

The following lemma on order of local conservation laws
for more general class of second-order evolution equations, which
covers class~\eqref{eqRDfghPowerG1}, is used below.

\begin{lemma}[\cite{Ivanova&Popovych&Sophocleous2006}]\label{LemmaOnOrderOfConsLawsOfDCEs}
Any local conservation law of any second-order $(1+1)$-dimensional
quasi-linear evolution equation has the first order and, moreover,
there exists its conserved vector with the density depending at
most on $t$, $x$, and $u$ and the flux depending at most on $t$,
$x$, $u$ and~$u_x$.
\end{lemma}

\begin{theorem}\label{TheoremOnClassificationCLsOfEqRDf1h}
A complete list of nonlinear equations of the form~(\ref{eqRDfghPowerG1}) having
nontrivial conservation laws is exhausted by the following ones
\setcounter{mcasenum}{0}

\vspace{1ex}

$\makebox[6mm][l]{\refstepcounter{mcasenum}\themcasenum\label{CaseCLsRDfghBinA}.}
m=n+1.$
\nopagebreak\\[1ex]\hspace*{3\parindent}%
$n\neq-1\colon\quad \bigl(\, \varphi^ifu,\,
-\varphi^iu^nu_x+\varphi^i_x\frac{u^{n+1}}{n+1}\,\bigr),\
\varphi^i,\ i=1,2.$

\vspace{1ex}

$\makebox[6mm][l]{\refstepcounter{mcasenum}\themcasenum\label{CaseCLsRDfghBinA1}.}
m=1,\ h=\mu f,\ \mu=\const.$
\nopagebreak\\[1ex]\hspace*{3\parindent}%
$n\neq-1 \colon\quad\bigl(\,xe^{-\mu t}fu,\ e^{-\mu t}\bigl(- x
u^nu_x+\frac{u^{n+1}}{n+1}\bigr)\,\bigr),\ xe^{-\mu t};$ \
$(\,e^{-\mu t}fu,\ - e^{-\mu t}u^nu_x\,),\ e^{-\mu t}.$
\nopagebreak\\[1ex]\hspace*{3\parindent}%
$n=-1\colon\quad (\,xe^{-\mu t}fu,\ e^{-\mu t}(-xu^{-1} u_x+\ln
u)\,),\ xe^{-\mu t};$ \ $(\,e^{-\mu t}fu,\ -e^{-\mu
t}u^{-1}u_x\,),\ e^{-\mu t}.$

\vspace{1ex}

$\makebox[6mm][l]{\refstepcounter{mcasenum}\themcasenum\label{CaseCLsRDfghBinA2}.}
m=0$.
\nopagebreak\\[1ex]\hspace*{3\parindent}%
$n\ne-1\colon\quad (\, xfu,\ -xu^nu_x+\frac{u^{n+1}}{n+1}-\int\! xh \, {\rm d}x\,),\ x;\ (\, fu,\
-u^nu_x-\int\! h \, {\rm d}x\,),\,1.$
\nopagebreak\\[1ex]\hspace*{3\parindent}%
$n=-1\colon\quad (\, xfu,\ -xu^{-1}u_x+\ln u-\int\! xh \, {\rm d}x\,),\ x;\ (\, fu,\
-u^{-1}u_x-\int\! h \, {\rm d}x\,),\,1.$

\vspace{1ex}

\noindent Here $\beta_1$ and $\beta_2$ are arbitrary constants.
The functions $\varphi^i=\varphi^i(x)$, $i=1,2$, form a
fundamental set of solutions of the second-order linear
ordinary differential equation $\varphi_{xx}+(n+1)h\varphi=0$.
(Together with constraints on the constant parameters $m$ and $n$
we also present conserved vectors and characteristics of the basis
elements of the corresponding space of conservation laws.)
\end{theorem}

\begin{proof}
In view of Lemma~\ref{LemmaOnOrderOfConsLawsOfDCEs}, we assume at once that $F=F(t,x,u)$ and $G=G(t,x,u,u_x)$. 
We substitute the expression for~$u_t$ deduced from~\eqref{eqRDfghPowerG1} into~(\ref{EqGen2DimConsLaw}) 
and split the obtained equation with respect to $u_{xx}$. 
The coefficient of~$u_{xx}$ gives the equation $u^n f^{-1}F_u+G_{u_x}=0$. 
Hence \[G=-u^nf^{-1}F_{u}u_x+\widehat G(t,x,u).\] 
Taking into account the derived expression for~$G$ and splitting the rest of equation~(\ref{EqGen2DimConsLaw}) 
with respect to the powers of $u_x$, we obtain the system of PDEs on the functions $F$ and $\widehat G$:
\begin{equation}\label{splitconslaw}
F_{uu}=0, \quad -u^n\bigg(\frac{F_{u}}f\bigg)_x+\widehat G_u=0,
\quad F_t+\frac hf u^m F_u+\widehat G_x=0.
\end{equation}
Solving the first two equations of~(\ref{splitconslaw}) yields
$F=\Phi(t,x)fu+F^0(t,x)$, $\widehat G=\Phi_x\int\! u^n{\rm d}u+G^0(t,x)$.
The case~$n=-1$ is special for the integration of the function~$u^n$ with respect to~$u$. So
\begin{gather*}
G=\begin{cases} 
-\Phi u^nu_x+\Phi_x\frac{u^{n+1}}{n+1}+G^0(t,x)& \mbox{if}\quad n\neq -1,
\\[0.5ex]
-\Phi u^{-1}u_x+\Phi_x\ln u+G^0(t,x) & \mbox{if}\quad n=-1.
\end{cases}
\end{gather*}
(It is convenient to single out~$f$ as a multiplier in the coefficient of~$u$ in the expression of~$F$.) 
Then the third equation of~(\ref{splitconslaw}) becomes 
\begin{equation*}\textstyle
f\Phi_tu+\Phi_{xx}\int\! u^n{\rm d}u+h\Phi u^{m}+F^0_t+G^0_x=0.
\end{equation*} 

In the further consideration the major role is played by the differential consequence of this equation obtained by differentiation with respect to~$u$,
\begin{equation}\label{EqRdfghClassifyingConditionForConsLaws}
f\Phi_t+\Phi_{xx}u^n+mh\Phi u^{m-1}=0.
\end{equation}
Indeed, it is the unique classifying condition for this problem, 
which should additionally be split with respect to~$u$ depending on values of the parameters~$n$ and~$m$.
It follows from~\eqref{EqRdfghClassifyingConditionForConsLaws}
that equation~\eqref{eqRDfghPowerG1} possesses nontrivial conserved vectors 
only for the special values of~$n$ and~$m$, namely, if
\[
(n,m)\in\{(n',0),\,(n',n'+1),\,(n',1),\, n'\in\mathbb R\}.
\]
We exclude the cases $(0,0)$ and $(0,1)$  from the consideration since
they correspond to linear equations of the form~\eqref{eqRDfghPowerG1}.
In all the classification cases except those with $m=0$ we obtain the equation
$F^0_t+G^0_x=0$ and hence we can assume up to equivalence of conserved vectors that then $F^0=G^0=0$.  
It is obvious that the conserved vector with $\Phi=0$ is trivial. 
Therefore, for nonzero conservation laws the corresponding value of the parameterizing function~$\Phi$ should also be nonzero.

\vspace{0.8ex}

1. $m=n+1$. Then equation~\eqref{EqRdfghClassifyingConditionForConsLaws} is split into two equations, $\Phi_t=0$ and $\Phi_{xx}+(n+1)h\Phi=0$, 
i.e., $\Phi$~depends only on $x$ and runs through the set of solutions of the second equation, 
which is a homogeneous linear second-order ordinary differential equation. $\Phi=C_1x+C_2$ if $n=-1$.

\vspace{0.8ex}

2. $m=1$, $n\neq0$. The splitting of~\eqref{EqRdfghClassifyingConditionForConsLaws} leads to the equations $\Phi_{xx}=0$ and $f\Phi_t+h\Phi=0$. 
We integrate the first equation and substitute the obtained expression $\Phi=\Phi^1(t)x+\Phi^2(t)$ for~$\Phi$ into the second equation.
Since $\Phi\neq0$, the resulting equation $\Phi^1_tfx+\Phi^2_tf+\Phi^1hx+\Phi^2h=0$ yields 
that $h=\mu f$ and $\Phi=C_3e^{-\mu t}x+C_4e^{-\mu t}$, where $\mu=\const$.

\vspace{0.8ex}

3. $m=0$, $n\neq0$. Equation~\eqref{EqRdfghClassifyingConditionForConsLaws} implies that $\Phi_{xx}=\Phi_t=0$, i.e., $\Phi=C_1x+C_2$.
Hence we have the equation $h\Phi+F^0_t+G^0_x=0$ for finding the functions $F^0$ and $G^0$. 
Its general solution is a sum of its particular solution $(F^0,G^0)=(0,-\int\!\Phi h\,{\rm d}x)$
and of the general solution of the associated homogeneous equation $F^0_t+G^0_x=0$. 
The second summand corresponds to a trivial conserved vector. 
\end{proof}

\begin{note}
We exclude the subcase $n=-1$ from the case $m=n+1$ of Theorem~8 since it coincides with the same subcase of the case $m=0$.
\end{note}

\begin{corollary}
For any (nonlinear) equation from class~\eqref{eqRDfghPower}, the
dimension of the space of local conservation laws is equal to
either 0 or 2. In the second case the equation can be reduced by a
point transformation to an equation from the same class, where
$g=1$ and $mh=0$. Then a~basis of the corresponding space of
characteristics is formed by the functions $\varphi^1=1$ and $\varphi^2=x$.
\end{corollary}

\section{Similarity solutions}\label{SectionOnRDfghSimilaritySolutions}

The Lie symmetry operators found as a result of solving the group classification problem
can be applied to construction of exact solutions of the corresponding equations.
The method of reductions with respect to subalgebras of Lie invariance algebras is
well-known and quite algorithmic to use in most cases; we refer to the standard textbooks
on the subject~\mbox{\cite{Olver1986,Ovsiannikov1982}}.
Exhaustive group analysis, including Lie reductions,
of nonlinear constant-coefficient diffusion--reaction equations with nonlinearities of general form
was carried out in~\cite{Dorodnitsyn1982} (see also~\cite{Ibragimov1994V1}).
A number of exact solutions of such equations are tabulated (see e.g.~\cite{Polyanin&Zaitsev}).
Then the problem on Lie reductions and Lie exact solutions of equations from class~\eqref{eqRDfghPower},
which have four- or five-dimensional Lie invariance algebras 
(or even three-dimensional Lie invariance algebras in case $m\not=1,n+1$), 
can be assumed as solved in view of Theorem~\ref{TheoremOnRDfghEqsWith4DimLieInvAlgs}.

%\looseness=-1
The Lie reduction algorithm is more effective for equations having
nice symmetry properties. Together with the above facts, it gives
solid argumentation for us to choose the cases from Table~1 with
at least one non-constant parameter-function of~$x$ and a
three-dimensional Lie invariance algebra. Namely, these are
Cases~9 and~12. (Let us remember that Case~5 can be transformed by
an additional equivalence transformation to the case with the same
value of~$f$ and~$\varepsilon=0$ and then reduced by equivalence
transformations to one of Cases~9--12 depending on the values
of~$n$ and~$f$.)

We have shown that equations
\[
u_t=(u^nu_x)_x+\alpha x^{-2}u^{n+1}
\qquad\mbox{and}\qquad
e^xu_t=(u^{-\frac 43}u_x)_x+\alpha u^{-\frac 13}
\]
(Table~1, Case~9 and Case~12) admit the three-dimensional Lie invariance algebras generated by
the operators
\[
X_1=\partial_t,\quad X_2=t\partial_t-\frac un\partial_u,\quad X_3=x\partial_x+\frac{2u}n\partial_u
\]
and
\[
X_1=\partial_t,\quad X_2=t\partial_t+\frac34u\partial_u,\quad X_3=\partial_x-\frac34u\partial_u
\]
correspondingly. Due to strong simplification of these cases with equivalence transformations,
the operators are simplified to simple linear combinations of translation and scale operators and,
therefore, are handy for further usage.
The both tuples of operators satisfy the commutation relations
\[
[X_1,X_2]=X_1, \quad [X_1,X_3]=0, \quad [X_2,X_3]=0.
\]
It means that the algebras are isomorphic to the algebra $\mathfrak{g}_2\oplus\mathfrak{g}_1$
being the direct sum of the two-dimensional non-Abelian Lie algebra~$\mathfrak{g}_2$ and
the one-dimensional Lie algebra~$\mathfrak{g}_1$.
An optimal set of subalgebras of~$\mathfrak{g}_2\oplus\mathfrak{g}_1$ can be easily constructed
with application of a standard technique~\mbox{\cite{Olver1986,Ovsiannikov1982}}.
Another way is to take the set from~\cite{Patera&Winternitz1977}.
(In this paper optimal sets of subalgebras are listed for all three- and four-dimensional algebras.)
The used optimal set consists~of
\begin{gather*}
\mbox{one-dimensional subalgebras:}\quad
\langle X_2 - \mu X_3\rangle, \quad \langle X_3\rangle, \quad
\langle X_1\pm X_3\rangle, \quad \langle X_1\rangle;\\
\mbox{two-dimensional subalgebras:}
\quad \langle X_1,X_3-\nu X_2\rangle,\quad \langle X_1, X_2\rangle,
\end{gather*}
where $\mu$ and $\nu$ are arbitrary constants.

Lie reduction to algebraic equations can be made only with the first two-dimensional subalgebra;
the second one does not satisfy the transversality condition~\cite{Olver1986}.
The corresponding ansatzes and reduced algebraic equations have the form
\begin{gather*}
9.\ u=Cx^\sigma,\quad \mbox{where}\quad \sigma=\frac{\nu+2}n;\quad C^{n+1}((n+1)\sigma^2-\sigma+\alpha)=0;
\\
12.\ u=Ce^{\sigma x},\quad \mbox{where}\quad \sigma=-\frac34(\nu+1);\quad C^{-\frac13}(\sigma^2-3\alpha)=0;
\end{gather*}
Here $C$ is an unknown constant to be found.
The reduced equations are compatible and have nontrivial (nonzero) solutions only for some values of~$\sigma$
and, moreover, become identities for these values of~$\sigma$.
As a result, the following stationary solutions are constructed:
\begin{gather*}
9.\ u=Cx^\sigma,\quad \mbox{where}\quad (n+1)\sigma^2-\sigma+\alpha=0;
\\
12.\ u=Ce^{\sigma x},\quad \mbox{where}\quad \sigma^2=3\alpha;
\end{gather*}
Here $C$ is an arbitrary constant.
These solutions can be also obtained with step-by-step reductions with respect to one-dimensional subalgebras.

The ansatzes and reduced equations corresponding to the one-dimensional subalgebras from the optimal system
are collected in table~2.

\begin{table}\footnotesize
\renewcommand{\arraystretch}{1.6}
%\caption{Lie symmetries}
\begin{center}
\textbf{Table 2.} Similarity reductions of Cases~9 and~12
\\[2ex]
\begin{tabular}{|r|c|c|c|c|}\hline
N\hfil\null& $X$ & $\omega$ & $u =$ & Reduced ODE \\
\hline
9.1 & $X_2 - \mu X_3$ & $xt^\mu$ &
$t^{-\frac{1+2\mu}{n}}\varphi(\omega)$
& $(\varphi^{n}{\varphi}_\omega)_\omega -\mu\omega\varphi_\omega+\frac{1+ 2\mu}{n}\varphi+
\alpha\omega^{-2}\varphi^{n+1}=0$
\\
9.2& $X_3$ & $t$ & $x^\frac 2n \varphi(\omega)$ &
 $n^2\varphi_\omega-(\alpha n^2+2n+4)\varphi^{n+1}=0\quad$
\\
9.3& $X_3\pm X_1$ & $xe^{\mp t}$ & $e^{\pm\frac {2t}n}\varphi(\omega)$ & $(\varphi^{n}{\varphi}_\omega)_\omega \pm\omega\varphi_\omega
\mp\frac 2n\varphi+\alpha\omega^{-2}\varphi^{n+1}=0$
\\
9.4& $X_1$&$x$&$\varphi(\omega)$&
$(\varphi^{n}{\varphi}_\omega)_\omega+\alpha\omega^{-2}\varphi^{n+1}=0$
\\
\hline
12.1 & $X_2 - \mu X_3$ & $x+\mu\ln{t}$ &
$t^{\frac 34(\mu+1)}\varphi(\omega)$
& $(\varphi^{-\frac 43}{\varphi}_\omega)_\omega -\mu e^\omega\varphi_\omega
-\frac 34 (\mu+1)e^\omega \varphi+\alpha\varphi^{-\frac 13}=0$
\\
12.2& $X_3$ & $t$ & $e^{-\frac 34 x} \varphi(\omega)$ &
$16\varphi_\omega+(3-16\alpha)\varphi^{-\frac 13}=0\quad$
\\
12.3& $X_3\pm X_1$ & $x\mp t$ & $e^{\mp\frac 34 t}\varphi(\omega)$ &
$(\varphi^{-\frac 43}{\varphi}_\omega)_\omega \pm e^\omega\varphi_\omega
\pm \frac 34 e^\omega \varphi+\alpha\varphi^{-\frac 13}=0$
\\
12.4& $X_1$&$x$&$\varphi(\omega)$&
$(\varphi^{-\frac 43}{\varphi}_\omega)_\omega+\alpha\varphi^{-\frac 13}=0$
\\
\hline
\end{tabular}
\end{center}
\end{table}

%\vspace{-2ex}

Some reduced equations are integrated completely:
\begin{gather*}
9.2.\ \varphi=\left(C-\Bigl(\alpha n+2+\frac4n\Bigr)\omega\right)^{-\frac1n};
\\[1ex]
9.4.\ \varphi=\begin{cases}
C_1\omega^\frac1{2(n+1)}(\ln{\omega}+C_2)^\frac1{n+1},
&\text{if}\quad \alpha'=0,\\[.5ex]
\bigl(C_1\omega^{\varkappa_1}+C_2\omega^{\varkappa_2}\bigr)^\frac1{n+1},
&\text{if}\quad \alpha'>0,\\[.5ex]
\omega^\frac1{2(n+1)}\bigl(C_1\sin(\sigma\ln{\omega})+C_2\cos(\sigma\ln{\omega})\bigr)^\frac 1{n+1},
&\text{if}\quad \alpha'<0,
\end{cases}
\\
\text{where}
\quad \alpha'=1-4\alpha(n+1),
\quad\varkappa_{1,2}=\frac{1\pm\sqrt{\alpha'}}2,
\quad \sigma=\frac{\sqrt{-\alpha'}}{2};
\end{gather*}
\vspace{-3ex}
\begin{gather*}
12.2.\ \varphi=\left(C+\left(\frac 43\alpha-\frac 14\right)\omega\right)^{\frac 34};
\\[1ex]
12.4.\ \varphi=\begin{cases}
C_1(\omega+C_2)^{-3},
&\text{if}\quad \alpha=0,\\[.5ex]
\bigl(C_1e^{\varkappa\omega}+C_2e^{-\varkappa\omega}\bigr)^{-3},
&\text{if}\quad \alpha>0,\\[.5ex]
\bigl(C_1\sin(\sigma\omega)+C_2\cos(\sigma\omega)\bigr)^{-3},
&\text{if}\quad \alpha<0,
\end{cases}
\\
\text{where}
\
\quad\varkappa=\sqrt{\alpha/3},
\quad \sigma=\sqrt{-\alpha/3}.
\end{gather*}
The above solutions of reduced equations results after substitution to the corresponding ansatzes to
exact solutions of the initial equations.

Let us note that the obtained solutions can by transformed by equivalence or admissible transformations to
exact solutions of much more complicated diffusion--reaction equations than the investigated ones.

\section{Conclusion}

In this paper different properties and objects concerning variable coefficient (1+1)-dimensional nonlinear
diffusion--reaction equations~\eqref{eqRDfghPower} are investigated in the framework of classical group analysis:
equivalence transformations, Lie symmetries, conservation laws and exact solutions.

The cornerstone of presented investigation is application of different generalizations of the usual equivalence group,
including generalized, extended and conditional equivalence groups and admissible transformations, in an optimal way via
gauging arbitrary elements.
It is the tool that makes principal simplifications in solving the group classification problem,
construction of conservation laws and finding exact solutions.
The set of admissible transformations of class~\eqref{eqRDfghPower} has an interesting structure
which is exhaustively described by Theorem~\ref{TheoremOnRDfghSetOfAdmTrans}.

It seems natural for one to consider other classes of PDEs in the
same way. The direct extension is investigation of  variable
coefficient (1+1)-dimensional nonlinear diffusion--reaction
equations for different (e.g. non-power) nonlinearities. Thus, we
will complete studying the Case~$n=0$ which is omitted in this
paper.

The adduced results form a basis for advanced analysis of
class~\eqref{eqRDfghPower} with `modern' symmetry methods. In
particular, classification of conservation laws creates the
necessary prerequisites for studying potential symmetries
\cite{bluman1989}. It is undoubted that generalized equivalences
will continue to play a singular role in these investigations.

\section*{Acknowledgements}
The research of R.\,P. was supported by Austrian Science Fund
(FWF), Lise Meitner project M923-N13. The research of O.\,V. was
partially supported by the grant of the President of Ukraine for
young scientists GF/F11/0061. A.\,G.\,J. would like to thank Eastern
University, Sri Lanka for its continuing support in his research
work. R.\,P., O.\,V. and A.\,G.\,J. are grateful for the hospitality
and financial support by the University of Cyprus.

\end{document}